\documentclass[letterpaper]{article} 
\usepackage{aaai23}  
\usepackage{times}  
\usepackage{helvet}  
\usepackage{courier}  
\usepackage[hyphens]{url}  
\usepackage{graphicx} 
\usepackage{natbib}  
\usepackage{caption} 
\usepackage{amsfonts,amsmath,amssymb,amsthm}
\usepackage{subcaption}
\usepackage{bm}
\usepackage{booktabs}

\DeclareMathOperator*{\argmin}{arg\,min}
\DeclareMathOperator*{\argmax}{arg\,max}
\newtheorem{definition}{Definition}
\newtheorem{theorem}{Theorem}
\newtheorem{lemma}{Lemma}
\newtheorem{fact}{Fact}

\newcommand{\norm}[1]{\Vert #1 \Vert}
\newcommand{\num}[1]{\vert #1 \vert}
\newcommand{\abso}[1]{\vert #1 \vert}
\newcommand{\ip}[1]{\langle #1 \rangle}

\makeatletter
\newif\if@restonecol
\makeatother

\usepackage[linesnumbered,ruled,vlined]{algorithm2e}
\usepackage{algpseudocode}

\SetKwComment{Comment}{$\triangleright$\ }{}

\title{SAH: Shifting-aware Asymmetric Hashing for Reverse $k$-Maximum Inner Product Search}
\author{
    Qiang~Huang,\textsuperscript{\rm 1}
    Yanhao~Wang,\textsuperscript{\rm 2}\thanks{Corresponding author}
    Anthony~K.~H.~Tung\textsuperscript{\rm 1}
}
\affiliations{
    \textsuperscript{\rm 1}School of Computing, National University of Singapore, Singapore\\
    \textsuperscript{\rm 2}School of Data Science and Engineering, East China Normal University, Shanghai, China\\
    huangq@comp.nus.edu.sg, yhwang@dase.ecnu.edu.cn, atung@comp.nus.edu.sg
}

\begin{document}

\maketitle

\begin{abstract}
This paper investigates a new yet challenging problem called Reverse $k$-Maximum Inner Product Search (R$k$MIPS).
Given a query (item) vector, a set of item vectors, and a set of user vectors, the problem of R$k$MIPS aims to find a set of user vectors whose inner products with the query vector are one of the $k$ largest among the query and item vectors.
We propose the first subquadratic-time algorithm, i.e., Shifting-aware Asymmetric Hashing (SAH), to tackle the R$k$MIPS problem.
To speed up the Maximum Inner Product Search (MIPS) on item vectors, we design a shifting-invariant asymmetric transformation and develop a novel sublinear-time Shifting-Aware Asymmetric Locality Sensitive Hashing (SA-ALSH) scheme. 
Furthermore, we devise a new blocking strategy based on the Cone-Tree to effectively prune user vectors (in a batch).
We prove that SAH achieves a theoretical guarantee for solving the RMIPS problem.
Experimental results on five real-world datasets show that SAH runs 4$\sim$8$\times$ faster than the state-of-the-art methods for R$k$MIPS while achieving F1-scores of over 90\%. The code is available at \url{https://github.com/HuangQiang/SAH}.
\end{abstract}

\section{Introduction}
\label{sec:intro}

Recommender systems based on Matrix Factorization~\cite{DBLP:journals/computer/KorenBV09} (MF) and Deep Matrix Factorization~\cite{DBLP:conf/ijcai/XueDZHC17} (DMF) models have been prevalent over the last two decades due to their precisely predictive accuracy, superior scalability, and high flexibility in various real-world scenarios.
In MF and DMF models, users and items are represented as vectors in a $d$-dimensional Euclidean space $\mathbb{R}^d$ obtained from a user-item rating matrix.
The relevance (or interestingness) of an item to a user is usually measured by the \emph{inner product} of their representing vectors.
This naturally gives rise to the Maximum Inner Product Search (MIPS) problem, which finds the vector in a set of $n$ item vectors $\mathcal{P} \subset \mathbb{R}^d$ that has the largest inner product with a query (user) vector $\bm{q} \in \mathbb{R}^d$, i.e., $\bm{p}^* = \argmax_{\bm{p} \in \mathcal{P}} \ip{\bm{p}, \bm{q}}$, as well as its extension $k$MIPS that finds $k$ ($k > 1$) vectors with the largest inner products for recommending items to users.
Due to its prominence in recommender systems, the $k$MIPS problem has attracted significant research interests, and numerous methods have been proposed to improve the search performance~\cite{DBLP:conf/kdd/RamG12,DBLP:conf/cikm/KoenigsteinRS12,DBLP:journals/ml/KeivaniSR18, DBLP:journals/tods/TeflioudiG17,DBLP:conf/sigmod/LiCYM17,DBLP:conf/icde/AbuzaidSBZ19, DBLP:conf/nips/Shrivastava014,DBLP:conf/icml/NeyshaburS15,DBLP:conf/uai/Shrivastava015,DBLP:conf/kdd/HuangMFFT18,DBLP:conf/nips/YanLDCC18, DBLP:conf/icdm/BallardKPS15,DBLP:conf/nips/YuHLD17,DBLP:conf/aistats/DingYH19,DBLP:conf/pkdd/LorenzenP20,DBLP:conf/kdd/Pham21, DBLP:conf/iccv/ShenLZYS15,DBLP:conf/aistats/GuoKCS16,DBLP:conf/aaai/DaiYNLC20,xiang2021gaips, DBLP:conf/nips/MorozovB18,DBLP:conf/emnlp/TanZXL19,DBLP:conf/nips/ZhouTX019,DBLP:conf/aaai/LiuYDLCY20,DBLP:conf/kdd/TanXZFZL21}. 

In this paper, we investigate a problem relevant to $k$MIPS yet less explored: \emph{how to find the users who are possibly interested in a given item?}
This problem is essential for market analysis from a \emph{reverse} perspective, i.e., the perspective of service providers instead of users.
For example, when an e-commerce service promotes a discounted product or launches a new product, a vital issue for designing an effective advertising campaign is identifying the customers who may want to buy this product.
In this case, the $k$MIPS might not be beneficial in finding potential customers: 
It can be leveraged to find $k$ user vectors having the largest inner products with the item vector, but these users might not be the target customers for the item if they are more interested in many other items than this item.
A more suitable formulation is to find the set of users for whom a query item is included in their $k$MIPS results, called Reverse $k$-Maximum Inner Product Search (R$k$MIPS). Formally,
\begin{definition}[R$k$MIPS~\cite{DBLP:conf/recsys/AmagataH21}]
Given an integer $k$ ($k \geq 1$), a query (item) vector $\bm{q} \in \mathbb{R}^d$, a set of $n$ item vectors $\mathcal{P} \subset \mathbb{R}^d$, and a set of $m$ user vectors $\mathcal{U} \subset \mathbb{R}^d$, the R$k$MIPS problem finds every user vector $\bm{u} \in \mathcal{U}$ such that $\bm{q}$ belongs to the $k$MIPS results of $\bm{u}$ among $\mathcal{P} \cup\{\bm{q}\}$.
\end{definition}

Compared with the $k$MIPS, the problem of R$k$MIPS is much more challenging.
The reasons are two folds.
First, the sizes of its search results vary among query vectors rather than being a fixed $k$. In the worst case, all $m$ users can be included in the R$k$MIPS results of a query $\bm{q}$.
Second, the number of items and users is typically large in real-world recommender systems.
A trivial approach is performing a linear scan over all items in $\mathcal{P} \cup \{\bm{q}\}$ for each user $\bm{u} \in \mathcal{U}$ and adding $\bm{u}$ to the R$k$MIPS results of $\bm{q}$ once $\bm{q}$ is included in the $k$MIPS results of $\bm{u}$.
For simplicity, we assume $m = O(n)$, i.e., $n$ and $m$ are of the same magnitude. This trivial approach takes $O(n^2 d)$ time, which is much higher than the time complexity of brute-force $k$MIPS and is often computationally prohibitive, especially for large $n$. 

Despite the importance of R$k$MIPS in real-world scenarios, little work has been devoted to studying this problem. Simpfer~\cite{DBLP:conf/recsys/AmagataH21} is a pioneer work yet the only known algorithm for solving R$k$MIPS. 
Its primary idea is to efficiently solve a decision version of $k$MIPS for each user. Simpfer maintains a lower-bound array of the $k$th largest inner product of size $k_{max}$ ($k \in \{1, 2, \cdots, k_{max}\}$) for each user $\bm{u}\in \mathcal{U}$ based on the $O(k_{max})$ items with the largest $l_2$-norms such that each user $\bm{u}$ can get a quick ``\emph{yes}''/``\emph{no}'' answer for any query $\bm{q}$ on whether it belongs to the $k$MIPS results of $\bm{u}$.
Moreover, it performs a linear scan using the Cauchy-Schwarz inequality to accelerate the $k$MIPS on item vectors. To reduce the number of user vectors for $k$MIPS, it partitions users into blocks based on their $l_2$-norms with a fixed-size interval.
Nonetheless, it is still a linear scan-based algorithm with the same worst-case time complexity of $O(n^2 d)$ as the trivial approach, and its performance degrades rapidly when $n$, $k$, or $d$ is large.

There have been many sublinear-time hashing schemes for solving approximate $k$MIPS~~\cite{DBLP:conf/nips/Shrivastava014,DBLP:conf/uai/Shrivastava015,DBLP:conf/icml/NeyshaburS15,DBLP:conf/kdd/HuangMFFT18,DBLP:conf/nips/YanLDCC18}. One can leverage these schemes to speed up the $k$MIPS for each $\bm{u} \in \mathcal{U}$. As such, the time to perform R$k$MIPS can be subquadratic. 
However, such an adaptation might still be less efficient and effective in practice. 
First, as $m$ is often larger than $n$, it is costly to check all users individually.
Second, there is no symmetric (or asymmetric) Locality-Sensitive Hashing (LSH) for MIPS in the original space $\mathbb{R}^d$~\cite{DBLP:conf/nips/Shrivastava014,DBLP:conf/icml/NeyshaburS15}. Existing hashing schemes develop different asymmetric transformations to convert MIPS into Nearest Neighbor Search (NNS) on angular (or Euclidean) distance, i.e., an item transformation $\bm{I}: \mathbb{R}^d \rightarrow \mathbb{R}^{d'}$ and a user transformation $\bm{U}: \mathbb{R}^d \rightarrow \mathbb{R}^{d'}$ on the item and user vectors, respectively, where $d' > d$. 
Unfortunately, these transformations add a large constant in angular (and Euclidean) distance, leading to a significant \emph{distortion error} for the subsequent NNS, i.e., the relative angular (and Euclidean) distance of any $\bm{I}(\bm{p})$ and $\bm{U}(\bm{u})$ will be much smaller than that in the original space. 
As a result, any $\bm{I}(\bm{p})$ can be the NNS result of $\bm{U}(\bm{u})$ even though their inner product $\ip{\bm{p},\bm{u}}$ is very small. Thus, the $k$MIPS results can be arbitrarily bad. 

In addition, the R$k$MIPS problem shares a similar concept with the reverse top-$k$ query~\cite{DBLP:conf/icde/VlachouDKN10,DBLP:journals/tkde/VlachouDKN11,DBLP:conf/sigmod/VlachouDNK13}, since both problems aim to find a set of users such that the query item is one of their top-$k$ results.
The methods for reverse top-$k$ queries, however, might not be suitable for solving R$k$MIPS as they usually assume that the dimensionality $d$ is low~\cite{DBLP:conf/sigmod/VlachouDNK13}, i.e., $d<10$, whereas $d$ is often dozens to hundreds in recommender systems. 
Another problem related to R$k$MIPS is the Reverse $k$-Nearest Neighbor Search (R$k$NNS)~\cite{DBLP:conf/sigmod/KornM00,DBLP:conf/icde/YangL01,DBLP:conf/cikm/SinghFT03,DBLP:conf/vldb/TaoPL04,DBLP:conf/sigmod/AchtertBKKPR06,DBLP:journals/corr/abs-1011-4955}.
Nevertheless, like the case of reverse top-$k$ queries, most existing R$k$NNS methods are also customized for low-dimensional data.

\paragraph{Our Contributions.}
In this paper, we propose the first subquadratic-time algorithm called Shifting-aware Asymmetric Hashing (SAH) to tackle the problem of R$k$MIPS in high-dimensional spaces. 
To accelerate the $k$MIPS on item vectors, we develop a provable, sublinear-time scheme called Shifting-Aware Asymmetric Locality-Sensitive Hashing (SA-ALSH) together with a novel shifting-invariant asymmetric transformation to reduce the distortion error significantly. 
Furthermore, we devise a novel blocking strategy for user vectors based on the Cone-Tree~\cite{DBLP:conf/kdd/RamG12}. Using the cone structure, we derive two tight upper bounds that can effectively prune user vectors (in a batch). 
SAH also inherits the basic idea of Simpfer~\cite{DBLP:conf/recsys/AmagataH21} to leverage the lower bounds for users to obtain a quick ``\emph{yes}''/``\emph{no}'' decision for the $k$MIPS. 
We prove that SAH achieves a theoretical guarantee for solving R$k$MIPS when $k=1$ in subquadratic time and space. 
In the experiments, we systematically compare SAH with a state-of-the-art $k$MIPS method H2-ALSH~\cite{DBLP:conf/kdd/HuangMFFT18} as well as the only known R$k$MIPS method Simpfer~\cite{DBLP:conf/recsys/AmagataH21}. Extensive results over five real-world datasets demonstrate that SAH runs 4$\sim$8$\times$ faster than them for R$k$MIPS while achieving F1-scores of over 90\%.

\section{Background}
\label{sec:back}
Before presenting SAH for solving R$k$MIPS, we first introduce the background of Locality-Sensitive Hashing (LSH) and Asymmetric Locality-Sensitive Hashing (ALSH). 

\subsection{Locality-Sensitive Hashing}
\label{sec:back:lsh}
LSH schemes are one of the most prevalent methods for solving high-dimensional NNS~\cite{indyk1998approximate,DBLP:conf/stoc/Charikar02,datar2004locality,andoni2006near,har2012approximate,andoni2015practical, huang2015query, lei2019sublinear, lei2020locality}. 
Given a hash function $h$, we say two vectors $\bm{p}$ and $\bm{u}$ collide in the same bucket if $h(\bm{p})=h(\bm{u})$.
Let $Dist(\bm{p},\bm{u})$ be a distance function of any two vectors $\bm{p}$ and $\bm{u}$.
Formally,
\begin{definition}[LSH Family~\cite{indyk1998approximate}]
\label{def:lsh-family}
Given a search radius $R$~$(R>0)$ and an approximation ratio $c$, a hash family $\mathcal{H}$ is called $(R, c R, p_1, p_2)$-sensitive to $Dist(\cdot,\cdot)$ if, for any $\bm{p}, \bm{u} \in \mathbb{R}^d$, it satisfies:
\begin{itemize}
\item If $Dist(\bm{p}, \bm{u}) \leq R$, then $\mathrm{Pr}_{h \in \mathcal{H}}[h((\bm{p}) = h(\bm{u})] \geq p_1$;
\item If $Dist(\bm{p}, \bm{u}) \geq c R$, then $\mathrm{Pr}_{h \in \mathcal{H}}[h(\bm{p}) = h(\bm{u})] \leq p_2$.
\end{itemize}
\end{definition}
An LSH family is valid for NNS only when $c > 1$ and $p_1 > p_2$.
With an $(R, c R, p_1, p_2)$-sensitive hash family, LSH schemes can deal with the NNS in sublinear time and subquadratic space.
\begin{theorem}[\citealt{indyk1998approximate}]
\label{theorem:time-complexity}
Given a family $\mathcal{H}$ of $(R, c R, p_1, p_2)$-sensitive hash functions, one can construct a data structure that finds an item vector $\bm{p}\in \mathcal{P}$ such that $Dist(\bm{p},\bm{u}) \leq c \cdot Dist(\bm{p^*},\bm{u})$ in $O(n^{1+\rho})$ space and $O(d n^\rho \log_{1/{p_2}}n)$ query time, where $\rho=\ln{p_1}/\ln{p_2}$ and $\bm{p^*} = \argmin_{\bm{p} \in \mathcal{P}} Dist(\bm{p}, \bm{u})$.
\end{theorem}

SimHash is a classic LSH scheme proposed by~\citet{DBLP:conf/stoc/Charikar02} for solving NNS on angular distance. The angular distance is computed as $\theta(\bm{p},\bm{u}) = \arccos(\frac{\ip{\bm{p},\bm{u}}}{\norm{\bm{p}} \norm{\bm{u}}})$ for any $\bm{p},\bm{u} \in \mathbb{R}^d$. Its LSH function is called Sign Random Projection (SRP), i.e.,
\begin{equation}
\label{eqn:angular-lsh-func}
h_{srp}(\bm{p}) = sgn(\ip{\bm{a},\bm{p}}), 
\end{equation}
where $\bm{a}$ is a $d$-dimensional vector with each entry drawn i.i.d.~from the standard normal distribution $\mathcal{N}(0,1)$;
$sgn(\cdot)$ and $\ip{\cdot,\cdot}$ denote the sign function and the inner product computation, respectively. 
Let $\delta = \theta(\bm{p},\bm{u})$ be the angular distance of any $\bm{p}$ and $\bm{u}$. The collision probability is:
\begin{equation}
\label{eqn:angular-col-prob}
p(\delta) = \Pr[h_{srp}(\bm{p})=h_{srp}(\bm{u})] = 1-\tfrac{\delta}{\pi}.
\end{equation}

\subsection{Asymmetric LSH}
\label{sec:back:alsh}

Since existing LSH schemes for NNS on Euclidean or angular distance are not directly applicable to MIPS, they usually perform asymmetric transformations to reduce MIPS to NNS, known as \emph{Asymmetric LSH} (ALSH).
\begin{definition}[ALSH Family~\cite{DBLP:conf/nips/Shrivastava014}]
\label{def:alsh-family}
Given an inner product threshold $S_0$ $(S_0 > 0)$ and an approximation ratio $c$, a hash family $\mathcal{H}$, along with two vector transformations, i.e., $\bm{I}: \mathbb{R}^d \rightarrow \mathbb{R}^{d'}$ (Item transformation) and $\bm{U}: \mathbb{R}^d \rightarrow \mathbb{R}^{d'}$ (User transformation), is called $(S_0, \frac{S_0}{c}, p_1, p_2)$-sensitive to the inner product $\ip{\cdot,\cdot}$ if, for any $\bm{p}, \bm{u} \in \mathbb{R}^d$, it satisfies:
\begin{itemize}
\item If $\ip{\bm{p}, \bm{u}} \geq S_0$, then $\mathrm{Pr}_{h \in \mathcal{H}}[h(\bm{I}(\bm{p})) = h(\bm{U}(\bm{u}))] \geq p_1$;
\item If $\ip{\bm{p}, \bm{u}} \leq \frac{S_0}{c}$, then $\mathrm{Pr}_{h \in \mathcal{H}}[h(\bm{I}(\bm{p})) = h(\bm{U}(\bm{u}))] \leq p_2$.
\end{itemize}
\end{definition}
The ALSH family is valid for MIPS only when $c > 1$ and $p_1 > p_2$. The transformation $\bm{I}$ ($\bm{U}$) is only applied to item vectors $\bm{p}\in \mathcal{P}$ (user vectors $\bm{u}\in \mathcal{U}$). The transformations are asymmetric if $\bm{I}(\bm{x}) \neq \bm{U}(\bm{x}) \neq \bm{x}$ for any $\bm{x} \in \mathbb{R}^d$. 

H2-ALSH~\cite{DBLP:conf/kdd/HuangMFFT18} is a state-of-the-art ALSH scheme for MIPS. Let $\bm{p} = [p_1, \cdots, p_d]$ and $\bm{u} = [u_1, \cdots, u_d]$. 
It designs a Query Normalized First (QNF) transformation to convert MIPS into NNS on Euclidean distance, which is defined as follows:
\begin{align}
\bm{I}(\bm{p}) & = [p_1, \cdots, p_d; \sqrt{M^2 - \norm{\bm{p}}^2}], 
\label{eqn:h2-item-trans}\\
\bm{U}(\bm{u}) & = [\lambda u_1, \cdots, \lambda u_d; 0],~\text{where}~\lambda = M/\norm{\bm{u}},
\label{eqn:h2-user-trans}
\end{align}
where $M$ is the maximum $l_2$-norm of all $\bm{p} \in \mathcal{P}$, i.e., $M = \max_{\bm{p} \in \mathcal{P}} \norm{\bm{p}}$, and $[\cdot;\cdot]$ denotes the concatenation of two vectors. Based on Equations~\ref{eqn:h2-item-trans} and~\ref{eqn:h2-user-trans}, we have
\begin{equation}
\label{eqn:h2-trans}
\norm{\bm{I}(\bm{p}) - \bm{U}(\bm{u})}^2 = 2M \cdot (M- \tfrac{\ip{\bm{p}, \bm{u}}}{\norm{\bm{u}}}).
\end{equation}

Let $\theta$ be the angle of $\bm{p}$ and $\bm{u}$. As $\ip{\bm{p}, \bm{u}}/{\norm{\bm{u}}} = \norm{\bm{p}}\cos\theta \leq \norm{\bm{p}} \leq M$, we have $M-\ip{\bm{p}, \bm{u}}/{\norm{\bm{u}}} \geq 0$. 
Since $M$ and $\norm{\bm{u}}$ are fixed, the MIPS in $\mathbb{R}^d$ can be converted into the NNS on Euclidean distance in $\mathbb{R}^{d+1}$. 
Unfortunately, the angle $\theta$ of any $\bm{p}$ and $\bm{u}$ in high-dimensional spaces is often close to $\pi/2$, incurring a very small $\cos\theta$. Thus, we often have $\norm{\bm{p}}\cos\theta \ll M$.
In the worst case, $\max_{\bm{p}\in \mathcal{P}} \norm{\bm{I}(\bm{p}) - \bm{U}(\bm{u})} / \min_{\bm{p}\in \mathcal{P}} \norm{\bm{I}(\bm{p}) - \bm{U}(\bm{u})} \rightarrow 1$. 
Suppose that this ratio is less than 2, and we set up $c=2$ for approximate NNS, which is a typical setting for LSH schemes \cite{tao2009quality, gan2012locality, huang2015query}.
Then, any $\bm{I}(\bm{p})$ can be the NNS result of $\bm{U}(\bm{u})$ even if $\ip{\bm{p}, \bm{u}}$ is small, which means that the MIPS result of H2-ALSH for $\bm{u}$ can be arbitrarily bad. 

H2-ALSH develops a homocentric hypersphere partition strategy to split the item vectors into different blocks with bounded $l_2$-norms such that the item vectors with smaller $l_2$-norms correspond to a smaller $M$. This strategy can alleviate the distortion error but still cannot remedy the issue caused by the angle close to $\pi/2$.

\section{The SAH Algorithm}
\label{sec:alg}

In this section, we propose the SAH algorithm to deal with the problem of R$k$MIPS on high-dimensional data. 

\subsection{Shifting-invariant Asymmetric Transformation}
\label{sec:alg:sat}

We first introduce a Shifting-invariant Asymmetric Transformation (SAT) that converts the MIPS in $\mathbb{R}^d$ into the NNS on angular distance in $\mathbb{R}^{d+1}$.
Let $\bm{c}$ be the centroid of the item set $\mathcal{P}$, i.e., $\bm{c} = [c_1,\cdots,c_d] = \frac{1}{n} \sum_{\bm{p} \in \mathcal{P}} \bm{p}$.
Suppose that $R$ is the radius of the smallest ball centered at $\bm{c}$ enclosing all $\bm{p} \in \mathcal{P}$, i.e., $R = \max_{\bm{p} \in \mathcal{P}} \norm{\bm{p}-\bm{c}}$.
Given any item vector $\bm{p}=[p_1,\cdots,p_d]$ and user vector $\bm{u}=[u_1,\cdots,u_d]$, the item transformation $\bm{I}: \mathbb{R}^d \rightarrow \mathbb{R}^{d+1}$ and user transformation $\bm{U}: \mathbb{R}^d \rightarrow \mathbb{R}^{d+1}$ of SAT are:
\begin{align}
  \bm{I}(\bm{p},\bm{c}) & = [p_1-c_1, \cdots, p_d-c_d; \sqrt{R^2 - \norm{\bm{p-c}}^2}], \label{eqn:sa-item-trans} \\
  \bm{U}(\bm{u}) & = [\lambda u_1, \cdots, \lambda u_d; 0], \text{where}~\lambda = R/\norm{\bm{u}}. \label{eqn:sa-user-trans}
\end{align}
As $\norm{\bm{I}(\bm{p},\bm{c})} = \norm{\bm{U}(\bm{u})} = R$, SAT maps each item vector $\bm{p} \in \mathcal{P}$ and user vector $\bm{u} \in \mathcal{U}$ in $\mathbb{R}^d$ to the hypersphere $\mathbb{S}^{d}$ of radius $R$. Based on Equations~\ref{eqn:sa-item-trans} and~\ref{eqn:sa-user-trans},
\begin{equation}
\label{eqn:sa-trans}
  \frac{\ip{\bm{I}(\bm{p},\bm{c}), \bm{U}(\bm{u})}}{\norm{\bm{I}(\bm{p},\bm{c})} \cdot  \norm{\bm{U}(\bm{u})}} = \frac{\ip{\bm{p}-\bm{c}, \bm{u}}}{R \cdot \norm{\bm{u}}}.
\end{equation}

The intuition of SAT comes from the fact that the MIPS result of any vector $\bm{u}$ is \emph{shift-invariant}, i.e., it is always the same no matter where the item vectors are shifted.
\begin{fact}
\label{fact:mips-shift-invariant}
Given a set of vectors $\mathcal{P}$, the MIPS result of any vector $\bm{u}$ is invariant whether all vectors in $\mathcal{P}$ are shifted by $\bm{c}$, i.e., $\argmax_{\bm{p}\in \mathcal{P}} \ip{\bm{p},\bm{u}} = \argmax_{\bm{p}\in \mathcal{P}} \ip{\bm{p}-\bm{c},\bm{u}}$.
\end{fact}

According to Fact~\ref{fact:mips-shift-invariant}, as the term $R \cdot \norm{\bm{u}}$ in Equation~\ref{eqn:sa-trans} is the same for any $\bm{p} \in \mathcal{P}$, we have $\argmax_{\bm{p}\in \mathcal{P}} \ip{\bm{p},\bm{u}} = \argmin_{\bm{p}\in \mathcal{P}} \arccos(\tfrac{\ip{\bm{I}(\bm{p},\bm{c}), \bm{U}(\bm{u})}}{\norm{\bm{I}(\bm{p},\bm{c})} \cdot \norm{\bm{U}(\bm{u})}})$.
Thus, SAT converts the MIPS in $\mathbb{R}^d$ into the NNS on angular distance in $\mathbb{R}^{d+1}$.
Compared with the QNF transformation, SAT introduces a shifting operation to the item vectors, which typically reduces the maximum $l_2$-norm among item vectors after the item transformation.
Moreover, according to Equation~\ref{eqn:sa-trans}, the distortion of SAT only comes from the ratio $R/\norm{\bm{p}-\bm{c}}$, which decreases the error caused by a small $\cos\theta$.
Therefore, SAT can significantly reduce the distortion error in practice, as will be validated in our experiments.

\subsection{Shifting-Aware Asymmetric LSH}
\label{sec:alg:saalsh}

\paragraph{ALSH Family.} 
We first describe the hash family $\mathcal{H}_{sa}$ of SA-ALSH for solving MIPS.
Let $h_{srp}(\cdot)$ be the SRP-LSH function in Equation~\ref{eqn:angular-lsh-func}.
Given the centroid $\bm{c}$ of item vectors, $\bm{I}(\bm{p},\bm{c})$ and $\bm{U}(\bm{u})$ in Equations~\ref{eqn:sa-item-trans} and~\ref{eqn:sa-user-trans}, respectively, the hash family $\mathcal{H}_{sa}$ of hash functions $h_{sa}$ is:
\begin{equation}
\label{eqn:sa-hash}
  h_{sa}(\bm{x}) =
  \begin{cases}
    h_{srp}(\bm{I}(\bm{x},\bm{c})), & \text{if}~\bm{x} \in \mathcal{P}, \\
    h_{srp}(\bm{U}(\bm{x})), & \text{if}~\bm{x} \in \mathcal{U}.
  \end{cases}
\end{equation}

According to Equation~\ref{eqn:sa-trans}, the collision probability of $h_{sa}(\cdot)$ for certain $\bm{p},\bm{u}$ is computed as follows: 
\begin{equation}
\label{eqn:sa-col-prob}
\Pr[h_{sa}(\bm{p}) = h_{sa}(\bm{u})] = p(\arccos(\tfrac{\ip{\bm{p}-\bm{c},\bm{u}}}{R \cdot \norm{\bm{u}}})),
\end{equation}
where $p(\cdot)$ is the collision probability of the SRP-LSH family as given in Equation~\ref{eqn:angular-col-prob}.
Let $p_1 = p(\arccos(\tfrac{S_0}{R \cdot \norm{\bm{u}}}))$ and $p_2 = p(\arccos(\tfrac{S_0}{c R \cdot \norm{\bm{u}}}))$. With $\bm{I}(\bm{p},\bm{c})$ and $\bm{U}(\bm{u})$, we show that $\mathcal{H}_{sa}$ is an ALSH family for $\ip{\cdot,\cdot}$.
\begin{lemma}
\label{lemma:sa-alsh-family}
Given an inner product threshold $S_0~(S_0>0)$, an approximation ratio $c~(c>1)$, and an $(\arccos(\tfrac{S_0}{R \cdot \norm{\bm{u}}}),$ $\arccos(\tfrac{S_0}{c R \cdot \norm{\bm{u}}}),p_1, p_2)$-sensitive SRP hash family $\mathcal{H}$ for the NNS on angular distance, the hash family $\mathcal{H}_{sa}$ of hash functions $h_{sa}(\cdot)$ is $(S_0, \frac{S_0}{c}, p_1, p_2)$-sensitive to $\ip{\cdot,\cdot}$.
\end{lemma}

\begin{proof}
According to Equation 8, we have
$\ip{\bm{p}-\bm{c},\bm{u}} \geq S_0 \Leftrightarrow \arccos(\tfrac{\ip{\bm{I}(\bm{p},\bm{c}), \bm{U}(\bm{u})}}{\norm{\bm{I}(\bm{p},\bm{c})} \cdot \norm{\bm{U}(\bm{u})}}) \leq \arccos(\tfrac{S_0}{R \cdot \norm{\bm{u}}})$ and $\ip{\bm{p}-\bm{c},\bm{u}} \leq \tfrac{S_0}{c} \Leftrightarrow \arccos(\tfrac{\ip{\bm{I}(\bm{p},\bm{c}), \bm{U}(\bm{u})}}{\norm{\bm{I}(\bm{p},\bm{c})} \cdot \norm{\bm{U}(\bm{u})}}) \geq \arccos(\tfrac{S_0}{c R \cdot \norm{\bm{u}}})$.

Recall that the collision probabilities are
$$p_1=p(\arccos(\tfrac{S_0}{R \cdot \norm{\bm{u}}})),\; p_2=p(\arccos(\tfrac{S_0}{c R \cdot \norm{\bm{u}}})).$$
Since the SRP hash family $\mathcal{H}$ is $\Big(\arccos(\tfrac{S_0}{R \cdot \norm{\bm{u}}}),\arccos$ $(\tfrac{S_0}{c R \cdot \norm{\bm{u}}}), p_1, p_2\Big)$-sensitive to the angular distance, we have
\begin{itemize}
\item $\ip{\bm{p}-\bm{c},\bm{u}} \geq S_0 \Rightarrow \Pr[h_{sa}(\bm{p}) = h_{sa}(\bm{u})] \geq p_1$;
\item $\ip{\bm{p}-\bm{c},\bm{u}} \leq \tfrac{S_0}{c} \Rightarrow \Pr[h_{sa}(\bm{p}) = h_{sa}(\bm{u})] \leq p_2$.
\end{itemize}
As $S_0>0$ and $c>1$, $\arccos(\tfrac{S_0}{c R \cdot \norm{\bm{u}}}) < \arccos(\tfrac{S_0}{R \cdot \norm{\bm{u}}})$.
According to Equation 2, we have $p_1 > p_2$. Thus, the hash family $\mathcal{H}_{sa}$ of hash functions $h_{sa}(\cdot)$ is $(S_0, \frac{S_0}{c}, p_1, p_2)$-sensitive to $\ip{\cdot,\cdot}$. 
\end{proof}

We now present SA-ALSH for performing MIPS with the newly designed ALSH family $\mathcal{H}_{sa}$. 
With the insight that the item vectors with larger $l_2$-norms belong to the MIPS results of user vectors with higher probability~\cite{DBLP:conf/kdd/HuangMFFT18,DBLP:conf/nips/YanLDCC18,DBLP:conf/aaai/LiuYDLCY20}, we introduce a data-dependent partitioning strategy to build the index separately for different norm-based partitions of item vectors.

\begin{algorithm}[t]
\caption{SA-ALSH Indexing}
\label{alg:sa-alsh-indexing}
\KwIn{A set of $n$ item vectors $\mathcal{P}$, an interval ratio $b \in (0, 1)$, number of hash tables $K \in \mathbb{Z}^+$;}
Compute $\norm{\bm{p}}$ for each item $\bm{p} \in \mathcal{P}$ and sort $\mathcal{P}$ in descending order of $\norm{\bm{p}}$\; \label{ln:sa-alsh-index:l2-and-sort}
$j = 0$;~$i = 0$\;
\While {$i < n$}{
  $j \gets j + 1$;~$M_j \gets \norm{\bm{p}_i}$;~$\mathcal{S}_j \gets \emptyset$\; \label{ln:sa-alsh-index:partition:start}
  \While {$i < n$ and $\norm{\bm{p}_i} > b M_j$} {
    $\mathcal{S}_j \gets \mathcal{S}_j \cup \{\bm{p}_i\}$;~$i \gets i + 1$\;
  } \label{ln:sa-alsh-index:partition:end}
  $\bm{c}_j = \frac{1}{\num{\mathcal{S}_j}} \sum_{\bm{p} \in \mathcal{S}_j} \bm{p}$;~$R_j=\max_{\bm{p} \in \mathcal{S}_j} \norm{\bm{p}-\bm{c}_j}$\; \label{ln:sa-alsh-index:centroid-and-radius}
  $\mathcal{I}_j \gets \emptyset$\; \label{ln:sa-alsh-index:item-trans:start}
  \ForEach{item $\bm{p} \in \mathcal{S}_j$}{
    $\bm{I}(\bm{p},\bm{c}_j) \gets [p_1-c_{j_1}, \cdots, p_d-c_{j_d}; \sqrt{R_j^2 - \norm{\bm{p} - \bm{c}_j}^2}]$\;
    $\mathcal{I}_j \gets \mathcal{I}_j \cup \{\bm{I}(\bm{p},\bm{c}_j)\}$\;
  } \label{ln:sa-alsh-index:item-trans:end}
  Build the $K$ hash tables for $\mathcal{I}_j$ using SimHash\; \label{ln:sa-alsh-index:simhash}
}
$t \gets j$\;
\end{algorithm}
\setlength{\textfloatsep}{1.5em} 

\paragraph{Indexing Phase.}
The indexing phase of SA-ALSH is depicted in Algorithm \ref{alg:sa-alsh-indexing}.
Given a set of item vectors $\mathcal{P}$, we first compute the $l_2$-norm $\norm{\bm{p}}$ for each $\bm{p} \in \mathcal{P}$ and sort them in descending order (Line~\ref{ln:sa-alsh-index:l2-and-sort}). 
Let $b$ be the interval ratio ($0 < b < 1$).
We partition $\mathcal{P}$ into $t$ disjoint subsets $\{\mathcal{S}_j\}_{j=1}^t$ such that $b M_j < \norm{\bm{p}} \leq M_j$ for each $\bm{p} \in \mathcal{S}_j$, where $M_j = \max_{\bm{p} \in \mathcal{S}_j} \norm{\bm{p}}$ (Lines~\ref{ln:sa-alsh-index:partition:start}--\ref{ln:sa-alsh-index:partition:end}).
For each $\mathcal{S}_j$, we first compute its centroid $\bm{c}_j$ and radius $R_j$, i.e., $\bm{c}_j = \frac{1}{\num{\mathcal{S}_j}} \sum_{\bm{p} \in \mathcal{S}_j} \bm{p}$ and $R_j=\max_{\bm{p} \in \mathcal{S}_j} \norm{\bm{p}-\bm{c}_j}$ (Line~\ref{ln:sa-alsh-index:centroid-and-radius}).
According to Equation~\ref{eqn:sa-item-trans}, we apply $\bm{I}: \mathbb{R}^d \rightarrow \mathbb{R}^{d+1}$ to convert each $\bm{p} \in \mathcal{S}_j$ into $\bm{I}(\bm{p}, \bm{c}_j)$ and generate a set of SRP-LSH functions to compute the hash values $h_{sa}(\bm{p})$ of each item $\bm{p}$ (Lines~\ref{ln:sa-alsh-index:item-trans:start}--\ref{ln:sa-alsh-index:item-trans:end}).
Finally, we apply SimHash to build the index for all $\bm{I}(\bm{p}, \bm{c}_j)$'s (Line~\ref{ln:sa-alsh-index:simhash}).

Note that $M_1, \cdots, M_t$ are sorted in descending order and thus can be leveraged to estimate the upper bound for pruning item vectors in a batch.
The number of partitions $t$ is \emph{automatically} determined by the $l_2$-norms of item vectors and the interval ratio $b$.
As the item vectors are partitioned into subsets and based on different centroids for the SAT, it is called the Shifting-Aware ALSH (SA-ALSH for short).

\paragraph{Query Phase.}
The query phase of SA-ALSH is shown in Algorithm \ref{alg:sa-alsh-mips}.
Given a user vector $\bm{u}$, SA-ALSH identifies a set of candidates $\mathcal{C}$ from $\mathcal{S}_1$ to $\mathcal{S}_t$ to determine whether the query $\bm{q}$ is included in the $k$MIPS result of $\bm{u}$. 
Let $\varphi$ be the largest inner product we find so far. We initialize $\mathcal{C}$ as an empty set (Line~\ref{ln:sa-alsh-mips:init}).
For each $\mathcal{S}_j$, we can determine an upper bound for the item vectors based on $M_j$ and $\norm{\bm{u}}$, i.e., $\mu_j = M_j \cdot \norm{\bm{u}}$ (Line~\ref{ln:sa-alsh-mips:upper-bound}). 
It is because based on the Cauchy-Schwarz inequality, as $b M_j < \norm{\bm{p}} \leq M_j$ for any $\bm{p} \in \mathcal{S}_j$, we have $\ip{\bm{p},\bm{u}} \leq \norm{\bm{p}} \cdot \norm{\bm{u}} \leq M_j \cdot \norm{\bm{u}}$.
If $\varphi > \mu_j$, we can prune $\mathcal{S}_j$ and the rest partitions $\{\mathcal{S}_{j+1}, \cdots, \mathcal{S}_t\}$ and return ``yes" (Line~\ref{ln:sa-alsh-mips:pruning}) because $M_j > M_{j+1}$ and $\mu_j > \mu_{j+1}$, $\bm{q}$ belongs to the $k$MIPS result of $\bm{u}$;
otherwise, we apply $\bm{U}: \mathbb{R}^d \rightarrow \mathbb{R}^{d+1}$ to convert $\bm{u}$ into $\bm{U}(\bm{u})$ (Line~\ref{ln:sa-alsh-mips:user-tranformation}), call SimHash for performing NNS on $\mathcal{S}_j$ (Line~\ref{ln:sa-alsh-mips:simhash}), and update $\varphi$ (Line~\ref{ln:sa-alsh-mips:update-varphi}).
If the inner product of $\bm{u}$ and $\bm{q}$ is smaller than $\varphi$, which means that $\bm{q}$ is not included in the $k$MIPS result of $\bm{u}$, we can safely stop and return ``no" (Line~\ref{ln:sa-alsh-mips:stop}).
Finally, we return ``yes" as $\bm{q}$ is kept in the $k$MIPS result of $\bm{u}$ (Line~\ref{ln:sa-alsh-mips:final}). 

\begin{algorithm}[t]
\caption{SA-ALSH} 
\label{alg:sa-alsh-mips}
\KwIn{User vector $\bm{u}$, query vector $\bm{q}$, $k \in \mathbb{Z}^+$;}
$\mathcal{C} = \emptyset$; $\varphi = -\infty$\; \label{ln:sa-alsh-mips:init}
\For{$j = 1$ to $t$}{
  $\mu_j = M_j \cdot \norm{\bm{u}}$\; \label{ln:sa-alsh-mips:upper-bound}
  \lIf{$\varphi > \mu_j$}{\Return{yes}} \label{ln:sa-alsh-mips:pruning}
  $\bm{U}(\bm{u}) = [\frac{R_j}{\norm{\bm{u}}} u_1, \cdots, \frac{R_j}{\norm{\bm{u}}}  u_d; 0]$\; \label{ln:sa-alsh-mips:user-tranformation}
  $\mathcal{C} \gets \mathcal{C}~\cup$~SimHash$(\mathcal{S}_j,\bm{U}(\bm{u})\big)$\; \label{ln:sa-alsh-mips:simhash}
  $\varphi \gets$~the $k$th largest inner product among the item vectors in $\mathcal{C}$ with $\bm{u}$\; \label{ln:sa-alsh-mips:update-varphi}
  \lIf{$\ip{\bm{u},\bm{q}} < \varphi$}{\Return{no}} \label{ln:sa-alsh-mips:stop}
}
\Return{yes}\; \label{ln:sa-alsh-mips:final}
\end{algorithm}

As SA-ALSH first performs $k$MIPS on the item vectors with the largest $l_2$-norms, which most probably contain the $k$MIPS result of $\bm{u}$, the search process can be stopped early and effectively avoid evaluating a large number of false positives.
Based on Theorem~\ref{theorem:time-complexity} and Lemma~\ref{lemma:sa-alsh-family}, we prove that SA-ALSH achieves a theoretical guarantee for solving MIPS in sublinear time and subquadratic space.
\begin{theorem}
\label{theo:sa-alsh-guarantee}
Given a hash family $\mathcal{H}_{SA}$ of hash functions $h_{sa}(\cdot)$ as defined by Equation~\ref{eqn:sa-hash}, SA-ALSH is a data structure which finds an item vector $\bm{p} \in \mathcal{P}$ for any user vector $\bm{u} \in \mathbb{R}^d$ such that $\ip{\bm{p},\bm{u}} \leq \ip{\bm{p^*},\bm{u}}/c$ with constant probability in $O(d n^{\rho}\log_{1/p_2}n)$ time and $O(n^{1+\rho})$ space, where $\rho = \ln (1/p_1) / \ln (1/p_2)$ and $\bm{p^*} = \arg\max_{\bm{p} \in \mathcal{P}} \ip{\bm{p},\bm{u}}$.
\end{theorem}

\begin{proof}
  According to Lemma~1, the hash family $\mathcal{H}_{sa}$ of hash functions $h_{sa}(\cdot)$ is an ALSH family which is locality-sensitive to the inner product $\ip{\cdot,\cdot}$.
  Given a set of item vectors $\mathcal{P}$ and any user vector $\bm{u} \in \mathbb{R}^d$, SA-ALSH leverages SAT to convert the MIPS into the NNS on angular distance, and it calls SimHash to find the NNS within an approximation ratio $c>1$.
  Therefore, according to Theorem~1, SA-ALSH is a data structure that can find an item vector $\bm{p} \in \mathcal{P}$ for any $\bm{u}$ such that $\ip{\bm{p},\bm{u}} \geq \ip{\bm{p}^*,\bm{u}}/c$ with constant probability in $O(d n^{\rho}\log_{1/p_2}n)$ time and $O(n^{1+\rho})$ space, where $\rho = \ln (1/p_1) / \ln (1/p_2)$ and $\bm{p}^* = \arg\max_{\bm{p} \in \mathcal{P}} \ip{\bm{p},\bm{u}}$.
\end{proof}

\subsection{Cone-Tree Blocking}
\label{sec:alg:cone}

Suppose that $\theta_{\bm{p},\bm{u}}$ is the angle of an item vector $\bm{p}$ and a user vector $\bm{u}$. As $\ip{\bm{p}, \bm{u}} = \norm{\bm{p}} \cdot \norm{\bm{u}} \cos\theta_{\bm{p},\bm{u}}$, we have another fact that the $l_2$-norm $\norm{\bm{u}}$ of $\bm{u}$ does not affect its MIPS result. Formally,
\begin{fact}
\label{fact:mips-independent-of-user-norm}
Given a set of item vectors $\mathcal{P}$, the MIPS result of any user vector $\bm{u}$ is independent of its $l_2$-norm $\norm{\bm{u}}$, i.e., $\argmax_{\bm{p} \in \mathcal{P}} \ip{\bm{p},\bm{u}} = \argmax_{\bm{p} \in \mathcal{P}} \norm{\bm{p}}\cos\theta_{\bm{p},\bm{u}}$.
\end{fact}

Based on Fact~\ref{fact:mips-independent-of-user-norm}, we simply assume that all user vectors are unit vectors, i.e., $\norm{\bm{u}} = 1$ for every $\bm{u} \in \mathcal{U}$.
Fact~\ref{fact:mips-independent-of-user-norm} also implies that the MIPS result is only affected by the direction of $\bm{u}$. Motivated by this, we design a new blocking strategy for user vectors based on Cone-Tree~\cite{DBLP:conf/kdd/RamG12}. 

\paragraph{Cone-Tree Structure.}
We first review the basic structure of Cone-Tree~\cite{DBLP:conf/kdd/RamG12}.
The Cone-Tree is a binary space partition tree.
Each node $N$ consists of a subset of user vectors, i.e., $N.S \subset \mathcal{U}$. Let $\num{N}$ be the number of user vectors in a node $N$, i.e., $\num{N} = \num{N.S}$. Any node $N$ and its two children $N.lc$ and $N.rc$ satisfy two properties: $\num{N.lc} + \num{N.rc} = \num{N}$ and $N.lc \cap N.rc = \emptyset$.
Specifically, $N.S = \mathcal{U}$ if $N$ is the root of the Cone-Tree. Each node maintains a cone structure for its user vectors, i.e., the center $N.\bm{c} = \tfrac{1}{\num{N}} \textstyle \sum_{\bm{u} \in N.S} \bm{u}$ and the maximum angle $N.\omega = \max_{\bm{u} \in N.S} \arccos(\tfrac{\ip{\bm{u}, N.\bm{c}}}{\norm{\bm{u}} \cdot \norm{N.\bm{c}}})$.

\paragraph{Upper Bounds for R$k$MIPS.}
Based on the cone structure, we present an upper bound to prune a group of user vectors for R$k$MIPS. Let $\phi$ be the angle of $N.\bm{c}$ and a query $\bm{q}$.
\begin{lemma}[Node-Level Upper Bound]
\label{lemma:node_upper_bound}
Let the function $\{\theta\}_+ = \max\{\theta,0\}$. Given a query $\bm{q}$ and a node $N$ that contains a subset of user vectors $N.S$ centered at $N.\bm{c}$ with the maximum angle $N.\omega$, the maximum possible $\ip{\bm{u},\bm{q}}$ of any user vector $\bm{u} \in N.S$ and $\bm{q}$ is bounded as follows:
\begin{equation}
\label{eqn:node_upper_bound}
\max_{\bm{u} \in N.S} \ip{\bm{u},\bm{q}} \leq \norm{\bm{q}} \cos(\{ \phi - N.\omega \}_+).
\end{equation}
\end{lemma}

\begin{proof}
  Recall that $\phi$ is the angle of the query $\bm{q}$ and the center $N.\bm{c}$, and we assume that the user vector $\norm{\bm{u}} = 1$ for every $\bm{u} \in N.S$.
  There are two cases for the relationship of $\phi$ and $N.\omega$ as follows:
  \begin{enumerate}
    \item ($\phi \leq N.\omega$). In this case, as the query $\bm{q}$ lies within the cone of $N$, $N.S$ might contain a user vector with the same direction as $\bm{q}$. Thus, $\max_{\bm{u} \in N.S} \ip{\bm{u},\bm{q}} \leq \norm{\bm{u}} \cdot \norm{\bm{q}} = \norm{\bm{q}}$;
    \item ($\phi > N.\omega$). In this case, we suppose that $\theta_{\bm{u},\bm{q}}$ is the angle of any $\bm{u} \in N.S$ and $\bm{q}$.
    We have
    \begin{align*}
    \max_{\bm{u} \in N.S} \ip{\bm{u},\bm{q}} \leq \norm{\bm{q}} \cdot \max_{\bm{u} \in N.S} \cos \theta_{\bm{u},\bm{q}}.
    \end{align*}
    Since $0 \leq \phi,N.\omega \leq \pi$ and $\phi > N.\omega$, $\theta_{\bm{u},\bm{q}}$ satisfies the inequality $0 < \phi - N.\omega \leq \theta_{\bm{u},\bm{q}} \leq \phi + N.\omega < 2 \pi$. 
    As $\cos\theta = \cos(2\pi-\theta)$ and $\cos \theta$ decreases monotonically as $\theta$ increases for any $\theta \in [0,\pi]$, the upper bound of $\cos \theta_{\bm{u},\bm{q}}$ is either $\cos(\phi - N.\omega)$ or $\cos(\phi + N.\omega)$. 
    As $0 \leq \phi, N.\omega \leq \pi$, we have $\sin \phi \geq 0$ and $\sin N.\omega \geq 0$ for any $\phi$ and $N.\omega$. Thus,  
    \begin{align*}
      \cos(\phi+N.\omega)
      & = \cos\phi \cos N.\omega - \sin\phi \sin N.\omega \\
      & \leq \cos\phi \cos N.\omega + \sin\phi \sin N.\omega \\
      & = \cos(\phi-N.\omega).
    \end{align*}
    Thus, $\max_{\bm{u} \in N.S} \ip{\bm{u},\bm{q}} \leq \norm{\bm{q}} \cdot \max_{\bm{u} \in N.S} \cos \theta_{\bm{u},\bm{q}} = \norm{\bm{q}} \cos(\phi - N.\omega)$.
  \end{enumerate}
  By combining the above two cases, we have $\max_{\bm{u} \in N.S} \ip{\bm{u},\bm{q}} \leq \norm{\bm{q}} \cos(\{ \phi - N.\omega \}_+)$ and the proof is concluded.
\end{proof}

The node-level upper bound can prune all user vectors within a node in a batch, whereas it might not be tight for each user vector.
To perform \emph{vector-level pruning}, we further maintain cone structures for the user vectors in each leaf node $N$. As such, we quickly get an upper bound for each $\bm{u} \in N.S$. The advantage is that all cones share the same center $N.\bm{c}$ and only an angle $\theta_{\bm{u}}$ of each $\bm{u}$ and $N.\bm{c}$ is kept.
\begin{lemma}[Vector-Level Upper Bound]
\label{lemma:point_upper_bound}
Given a query $\bm{q}$ and a leaf node $N$ that maintains the angle $\theta_{\bm{u}}$ of each user vector $\bm{u} \in N.S$ and the center $N.\bm{c}$, the maximum possible $\ip{\bm{u},\bm{q}}$ of each $\bm{u} \in N.S$ and $\bm{q}$ is bounded as follows:
\begin{equation}
\label{eqn:point_upper_bound}
\ip{\bm{u},\bm{q}} \leq \norm{\bm{q}} \cos(\abso{\phi - \theta_{\bm{u}}}).
\end{equation}
\end{lemma}

\begin{proof}
  Recall that $\phi$ is the angle of the query $\bm{q}$ and the center $N.\bm{c}$, and $\theta_{\bm{u}}$ is the angle of the user vector $\bm{u} \in N.S$ and $N.\bm{c}$. Suppose $\theta_{\bm{u},\bm{q}}$ is the angle of $\bm{u}$ and $\bm{q}$. As $0 \leq \phi,\theta_{\bm{u}} \leq \pi$, based on the triangle inequality, we have $0 \leq \abso{\phi-\theta_{\bm{u}}} \leq \theta_{\bm{u},\bm{q}} \leq \phi+\theta_{\bm{u}} \leq 2\pi$. As $\norm{\bm{u}} = 1$, we have
  \begin{align*}
    \ip{\bm{u},\bm{q}} 
    &= \norm{\bm{u}} \cdot \norm{\bm{q}} \cos \theta_{\bm{u},\bm{q}} \\
    &\leq \norm{\bm{q}} \cdot \max_{\theta_{\bm{u},\bm{q}} \in [\abso{\phi-\theta_{\bm{u}}}, \phi+\theta_{\bm{u}}]} \cos\theta_{\bm{u},\bm{q}}. 
  \end{align*}
  As $\cos\theta = \cos(2\pi-\theta)$ and $\cos \theta$ decreases monotonically as $\theta$ increases for any $\theta \in [0,\pi]$, the upper bound of $\cos\theta_{\bm{u},\bm{q}}$ is either $\cos(\abso{\phi-\theta_{\bm{u}}})$ or $\cos(\phi+\theta_{\bm{u}})$. The relationship of $\phi$ and $\theta_{\bm{u}}$ also consists of the following two cases:
  \begin{enumerate}
    \item ($\phi \geq \theta_{\bm{u}}$). In this case, $\theta_{\bm{u},\bm{q}}$ satisfies the inequality $0 \leq \phi - \theta_{\bm{u}} \leq \theta_{\bm{u},\bm{q}} \leq \phi + \theta_{\bm{u}} \leq 2\pi$. As $0 \leq \phi,\theta_{\bm{u}} \leq \pi$, we have $\sin \phi \geq 0$ and $\sin \theta_{\bm{u}} \geq 0$ for any $\phi$ and $\theta_{\bm{u}}$. Thus,
    \begin{align*}
      \cos(\phi+\theta_{\bm{u}})
      & = \cos\phi \cos \theta_{\bm{u}} - \sin\phi \sin \theta_{\bm{u}} \\
      & \leq \cos\phi \cos \theta_{\bm{u}} + \sin\phi \sin \theta_{\bm{u}} \\
      & = \cos(\phi-\theta_{\bm{u}}).
    \end{align*}
    Thus, we have $\ip{\bm{u},\bm{q}} \leq \norm{\bm{q}} \cdot \max_{\theta_{\bm{u},\bm{q}} \in [\phi-\theta_{\bm{u}}, \phi+\theta_{\bm{u}}]} \cos\theta_{\bm{u},\bm{q}} = \norm{\bm{q}} \cdot \cos(\phi-\theta_{\bm{u}})$.
    \item ($\phi < \theta_{\bm{u}}$). In this case, $\theta_{\bm{u},\bm{q}}$ satisfies the inequality $0 < \theta_{\bm{u}}-\phi \leq \theta_{\bm{u},\bm{q}} \leq \theta_{\bm{u}}+\phi < 2\pi$. Similar to Case~(1), we can prove that $\cos(\theta_{\bm{u}}+\phi) \leq \cos(\theta_{\bm{u}}-\phi)$. Thus, $\ip{\bm{u},\bm{q}} \leq \norm{\bm{q}} \cdot \max_{\theta_{\bm{u},\bm{q}} \in [\theta_{\bm{u}}-\phi, \theta_{\bm{u}}+\phi]} \cos\theta_{\bm{u},\bm{q}} = \norm{\bm{q}} \cos(\theta_{\bm{u}}-\phi)$.
  \end{enumerate}
  By combining the above two cases, we have $\ip{\bm{u},\bm{q}} \leq \norm{\bm{q}} \cos(\abso{\phi - \theta_{\bm{u}}})$ and conclude the proof.
\end{proof}

Compared with the Cauchy-Schwarz inequality $\ip{\bm{u},\bm{q}} \leq \norm{\bm{u}} \cdot  \norm{\bm{q}} = \norm{\bm{q}}$ used in Simpfer~\cite{DBLP:conf/recsys/AmagataH21}, since $\cos(\abso{\phi - \theta_{\bm{u}}}) \leq 1$, Equation~\ref{eqn:point_upper_bound} is strictly tighter.

\paragraph{Cone-Tree Blocking.}
We now present our blocking strategy based on Cone-Tree to partition the set of user vectors $\mathcal{U}$.
We first show the pseudocode of Cone-Tree construction in Algorithm \ref{alg:cone-construct}. The center $N.\bm{c}$ and the maximum angle $N.\omega$ are maintained within each node $N$ (Lines~\ref{ln:cone:center}~\&~\ref{ln:cone:max-angle}).
For an internal node $N$, the splitting procedure is performed with three steps: (1) we select a random point $\bm{v} \in N.S$ (Line~\ref{ln:cone:random-point}); (2) we find the point $\bm{u}_l$ with the minimum inner product of $\bm{v}$, i.e., $\bm{u}_l = \argmin_{\bm{u} \in S} \ip{\bm{u},\bm{v}}$ (Line~\ref{ln:cone:pivot1}); (3) we find another point $\bm{u}_r$ with the minimum inner product of $\bm{u}_l$, i.e., $\bm{u}_r = \argmin_{\bm{u} \in S} \ip{\bm{u},\bm{u}_l}$ (Line~\ref{ln:cone:pivot2}). 
As we assume that all user vectors are unit vectors, the point $\bm{u}_r$ with the minimum inner product of $\bm{u}_l$ is also the one with the largest angle. As such, we use linear time (i.e., $O(\num{N} \cdot d)$) to efficiently find a pair of pivot vectors $\bm{u}_l$ and $\bm{u}_r$ with a large angle.
Then, we assign each $\bm{u} \in N.S$ to the pivot having a smaller angle with $\bm{u}$ and thus split $S$ into two subsets $S_l$ and $S_r$ accordingly (Lines~\ref{ln:cone:left-set}~\&~\ref{ln:cone:right-set}). 
For a leaf node $N$, we additionally maintain the angle $\theta_{\bm{u}}$ between each $\bm{u} \in N.S$ and $N.\bm{c}$ (Lines~\ref{ln:cone:theta-start}~\&~\ref{ln:cone:theta-end}). 
Suppose $N_0$ is the maximum leaf size. We start by assigning all user vectors in $\mathcal{U}$ to the root node. The Cone-Tree is built by performing the splitting procedure recursively from the root node until all leaf nodes contain at most $N_0$ user vectors. 
By the Cone-Tree construction, each leaf node contains a set of user vectors close to each other.
Thus, the leaf nodes are used as the blocks of user vectors.

\begin{algorithm}[t]
\caption{Cone-Tree Construction}
\label{alg:cone-construct}
\KwIn{Subset $S \subseteq \mathcal{U}$, maximum leaf size $N_0$;}
$N.S \gets S$;~$N.\bm{c} \gets \frac{1}{\num{N}} \sum_{\bm{u} \in N.S} \bm{u}$\; \label{ln:cone:center}
$N.\omega \gets \max_{\bm{u} \in N.S} \arccos(\tfrac{\ip{\bm{u},N.\bm{c}}}{\norm{\bm{u}} \cdot \norm{N.\bm{c}}})$\; \label{ln:cone:max-angle}
\uIf(\Comment*[f]{internal node}){$\num{N} > N_0$}{
  Select a point $\bm{v} \in N.S$ uniformly at random\; \label{ln:cone:random-point}
  $\bm{u}_l \gets \arg\min_{\bm{u} \in S} \ip{\bm{u},\bm{v}}$\; \label{ln:cone:pivot1}
  $\bm{u}_r \gets \arg\min_{\bm{u} \in S} \ip{\bm{u},\bm{u}_l}$\; \label{ln:cone:pivot2}
  $S_l \gets \{\bm{u} \in S \mid \cos \theta_{\bm{u},\bm{u}_l} \geq \cos \theta_{\bm{u},\bm{u}_r}\}$\; \label{ln:cone:left-set}
  $S_r \gets S \setminus S_l$\; \label{ln:cone:right-set}
  $N.lc \gets $~Cone-Tree Construction($S_l, N_0$)\;
  $N.rc \gets $~Cone-Tree Construction($S_r, N_0$)\;
  \Return $N$\;
}
\Else(\Comment*[f]{leaf node}){
  \ForEach{$\bm{u} \in N.S$} { \label{ln:cone:theta-start}
    $\theta_{\bm{u}} \gets \arccos(\tfrac{\ip{\bm{u},N.\bm{c}}}{\norm{\bm{u}} \cdot \norm{N.\bm{c}}})$\; 
  } \label{ln:cone:theta-end}
  \Return $N$\;
}
\end{algorithm}

\subsection{Shifting-aware Asymmetric Hashing}
\label{sec:alg:sah}
Finally, we combine SA-ALSH with the Cone-Tree blocking strategy and propose the Shifting-aware Asymmetric Hashing (SAH) algorithm for solving R$k$MIPS.
SAH uses SA-ALSH to speed up the $k$MIPS on item vectors.
Furthermore, it leverages the Cone-Tree blocking strategy along with node and vector-level upper bounds for pruning user vectors.
In addition, it inherits the basic idea of Simpfer~\cite{DBLP:conf/recsys/AmagataH21} to utilize the lower-bound arrays of user vectors to get a quick answer for the $k$MIPS.  

\begin{algorithm}[t]
\caption{SAH Indexing}
\label{alg:sah-indexing}
\KwIn{Item set $\mathcal{P}$, user set $\mathcal{U}$, $k_{max} \in \mathbb{Z}^+$, maximum leaf size $N_0$;}
Compute $\norm{\bm{p}_j}$~for~each~item~$\bm{p}_j \in \mathcal{P}$\;\label{ln:index:l2}
Sort $\mathcal{P}$ in descending order of $\norm{\bm{p}_j}$\;\label{ln:index:sort}
$\mathcal{P}^\prime \gets$~the first $O(k_{max})$ items of $\mathcal{P}$\;\label{ln:index:lb:s}
\ForEach{user $\bm{u}_i \in \mathcal{U}$}{
  $\mathcal{S} \gets$~$k_{max}$MIPS of $\bm{u}_i$ on $\mathcal{P}^\prime$\;
  \ForEach{$\bm{p}^*_j \in \mathcal{S}$}{
     $L_i^j \gets \ip{\bm{u}_i,\bm{p}^*_j}$\;
  }
  \label{ln:index:lb:t}
}
Build an SA-ALSH index for $\mathcal{P} \setminus \mathcal{P}^\prime$\;\label{ln:index:sa-alsh-index}
$T \gets$~Cone-Tree Construction($\mathcal{U}$, $N_0$)\; \label{ln:index:cone-tree-construct}
$\mathcal{B} \gets$~Extract all leaf nodes from $T$\; \label{ln:index:cone-tree-extract}
\ForEach{block $B \in \mathcal{B}$\label{ln:index:blk:s}}{
  \ForEach{user $\bm{u}_i \in B$}{
    \For{$j=1$ \KwTo $k_{max}$}{
      $L^j(B) \gets \min\{L^j(B), L_i^j\}$\;
    }
  }
  \label{ln:index:blk:t}
}
\end{algorithm}

\paragraph{Indexing Phase.}
The indexing phase of SAH is depicted in Algorithm~\ref{alg:sah-indexing}.
Given a set of item vectors $\mathcal{P}$ and a set of user vectors $\mathcal{U}$, it first computes $\norm{\bm{p}_j}$ for each $\bm{p}_j \in \mathcal{P}$ and sort $\mathcal{P}$ in descending order of $\norm{\bm{p}_j}$ (Lines~\ref{ln:index:l2}~\&~\ref{ln:index:sort}).
Let $\mathcal{P}^\prime$ be the first $O(k_{max})$ item vectors in the sorted $\mathcal{P}$. 
It then retrieves the $k_{max}$MIPS results $\mathcal{S}$ on $\mathcal{P}^\prime$ and computes a lower-bound array $L_i$ of size $k_{max}$ for each $\bm{u}_i \in \mathcal{U}$, i.e., $L_i^j = \ip{\bm{u}_i,\bm{p}_j^*}$, where $\bm{p}_j^*$ is the item of the $j$th ($1 \leq j \leq k_{max}$) largest inner product in $\mathcal{S}$ (Lines~\ref{ln:index:lb:s}--\ref{ln:index:lb:t}).
Next, it calls Algorithm \ref{alg:sa-alsh-indexing} to builds an SA-ALSH index for $\mathcal{P} \setminus \mathcal{P}^\prime$ (Line~\ref{ln:index:sa-alsh-index}) and calls Algorithm \ref{alg:cone-construct} to build a Cone-Tree $T$ for $\mathcal{U}$ (Line~\ref{ln:index:cone-tree-construct}).
Then, it extracts each leaf node in the Cone-Tree $T$
as a block $B \in \mathcal{B}$ for SAH (Line~\ref{ln:index:cone-tree-extract}).
Finally, it maintains a lower-bound array $L(B)$ of size $k_{max}$ for each block $B \in \mathcal{B}$ (Lines~\ref{ln:index:blk:s}--\ref{ln:index:blk:t}).

\paragraph{Query Phase.}
The query phase of SAH is shown in Algorithm~\ref{alg:sah-rkmips}.
Since every user vector $\bm{u}_i$ might be included in the R$k$MIPS results of the query $\bm{q}$, SAH checks each block $B \in \bm{B}$ individually.
According to Lemma~\ref{lemma:node_upper_bound}, it first verifies if $\norm{\bm{q}} \cos(\{\phi - N.\omega\}_+) < L^k(B)$ or not (Line~\ref{ln:query:lb:node}). If yes, it will be safe to skip all user vectors in $B$; otherwise, each user $\bm{u}_i$ in $B$ should be further checked. 
Then, based on Lemma~\ref{lemma:point_upper_bound}, it determines whether $\norm{\bm{q}} \cos(\abso{\phi - \theta_{\bm{u_i}}}) < L_i^k$ (Line~\ref{ln:query:lb:user}). If yes, $\bm{u}_i$ will be pruned; otherwise, it computes the actual inner product $\ip{\bm{u}_i,\bm{q}}$ and prunes $\bm{u}_i$ when $\ip{\bm{u}_i,\bm{q}} < L_i^k$ (Line~\ref{ln:query:ip:user}).
If $\bm{u}_i$ still cannot be pruned, since the item vectors are sorted in descending order of their $l_2$-norms, $\norm{\bm{u}_i} \cdot \norm{\bm{p}_k} = \norm{\bm{p}_k}$ is the upper bound of the $k$th largest inner product of $\bm{u}_i$. If $\ip{\bm{u}_i,\bm{q}} \geq \norm{\bm{p}_k}$, $\bm{u_i}$ will be added to the R$k$MIPS results $\mathcal{S}$ of $\bm{q}$; otherwise, SA-ALSH will be used for performing $k$MIPS on $\bm{u}_i$ to decide whether $\bm{q}$ is in the $k$MIPS results among $\mathcal{P} \cup \{\bm{q}\}$, and $\bm{u}_i$ will be added to $\mathcal{S}$ if yes (Lines~\ref{ln:query:alsh:s}--\ref{ln:query:alsh:t}).
Finally, it returns $\mathcal{S}$ as the R$k$MIPS results of $\bm{q}$ (Line~\ref{ln:query:sol}).

Based on Theorem~\ref{theo:sa-alsh-guarantee}, we show that SAH has a theoretical guarantee for RMIPS in subquadratic time and space.
\begin{theorem}
\label{theo:sah-guarantee}
Given a set of $n$ item vectors $\mathcal{P}$ and a set of $m$ user vectors $\mathcal{U}$ ($m=O(n)$), SAH is a data structure that finds each user vector $\bm{u} \in \mathcal{U}$ for any query vector $\bm{q} \in \mathbb{R}^d$ such that $\bm{q}$ is the MIPS result of $\bm{u}$ among $\mathcal{P} \cup\{\bm{q}\}$ with constant probability in $O(d n^{1+\rho} \log_{1/p_2}{n})$ time and $O(n^{1+\rho})$ space, where $\rho = \ln (1/p_1) / \ln (1/p_2)$.
\end{theorem}

\begin{proof}
  First, the space usage of SAH consists of the following three parts.
  \begin{itemize}
    \item SAH stores the hash tables of SA-ALSH for $t$ disjoint subsets $\{\mathcal{S}_1, \mathcal{S}_2, \cdots, \mathcal{S}_t\}$, where $\sum_{i=1}^t \num{\mathcal{S}_i} = n$. According to Theorem 2, SA-ALSH takes $O({\num{\mathcal{S}_i}}^{1+\rho})$ space for each $\mathcal{S}_i$, where $\rho=\ln(1/p_1)/\ln(1/p_2)<1$. As $\num{\mathcal{S}_i} \leq n$, we have $\sum_{i=1}^t \num{\mathcal{S}_i}^{1+\rho} \leq \sum_{i=1}^t \num{\mathcal{S}_i} \cdot n^{\rho} = n^{1+\rho}$. Thus, the hash tables take $O(n^{1+\rho})$ space.
    \item SAH stores the blocks of user vectors, i.e., the leaf nodes of Cone-Tree. Suppose that there are $l$ leaf nodes in Cone-Tree. For each leaf node $N_j$, SAH takes $O(d)$ space to store the center, $O(\num{N_j})$ space to store the angle $\theta_{\bm{u}}$ of each user vector $\bm{u} \in {N_j}.S$ and $O(k_{max})$ space to store the lower bounds of this block. Since $\sum_{j=1}^{l} \num{N_j}=m$, $l \ll m$, and $k_{max}=O(1)$, the blocks takes $\sum_{j=1}^{l} (d+\num{N_j}+k_{max}) = m + l \cdot (d+k_{max}) = O(m)$ space in total.
    \item SAH takes $O(m \cdot k_{max})$ space to store the lower bounds of each user vector.
  \end{itemize}
  Suppose that $m=O(n)$ and $k_{max} = O(1)$, the space complexity of SAH is thus $O(n^{1+\rho} + n + m \cdot k_{max}) = O(n^{1+\rho})$.
  
  To answer an RMIPS query, SAH requires performing MIPS for each user vector $\bm{u} \in \mathcal{U}$ in the worst case.
  According to Theorem 2, SA-ALSH finds an (approximate) MIPS result of $\bm{u}$ among $\mathcal{P} \cup \{\bm{q}\}$ with constant probability in $O(dn^{\rho}\log_{1/p_2} n)$ time.
  Since $m = O(n)$, the time complexity of performing MIPS using SAH is $O(m \cdot dn^{\rho}\log_{1/p_2} n) = O(dn^{1+\rho}\log_{1/p_2} n)$.
\end{proof}

\begin{algorithm}[t]
\caption{SAH} 
\label{alg:sah-rkmips}
\KwIn{query vector $\bm{q}$, $k \in \mathbb{Z}^+$, item set $\mathcal{P}$, user set $\mathcal{U}$, Cone-Tree blocks $\mathcal{B}$;}
$\mathcal{S} \gets \emptyset$\;
\ForEach{block $B \in \mathcal{B}$}{
  \textbf{if}~$\norm{\bm{q}} \cos(\{ \phi - N.\omega \}_+) < L^k(B)$~\textbf{then}~\textbf{continue}\label{ln:query:lb:node}\;
    \ForEach{user $\bm{u}_i \in B$}{
      \lIf{$\norm{\bm{q}} \cos(\abso{\phi - \theta_{\bm{u}_i}}) < L_i^k$\label{ln:query:lb:user}}{\textbf{continue}}
      \textbf{if}~$\ip{\bm{u}_i,\bm{q}} < L_i^k$~\textbf{then}~\textbf{continue}\;\label{ln:query:ip:user}
      \uIf{$\norm{\bm{p}_k} > \ip{\bm{u}_i,\bm{q}}$\label{ln:query:alsh:s}}{
        $ans~\gets$~SA-ALSH$(\bm{u}_i, \bm{q}, k)$\;
        \textbf{if}~$ans = yes$~\textbf{then}~$\mathcal{S} \gets \mathcal{S} \cup \{\bm{u}_i\}$\;
      }
      \Else{
        $\mathcal{S} \gets \mathcal{S} \cup \{\bm{u}_i\}$\;
      }
      \label{ln:query:alsh:t}
    }
}
\Return $\mathcal{S}$\;\label{ln:query:sol}
\end{algorithm}

\section{Experiments}
\label{sec:exp}

\begin{figure*}[t]
  \captionsetup{skip=3pt}
  \captionsetup[sub]{labelformat=empty,skip=0pt,position=top,font={scriptsize}}
  \centering
  \includegraphics[height=0.11in]{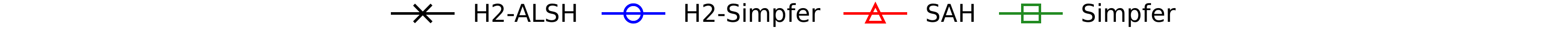}
  \\ \vspace{-1em}
  \subcaptionbox{}[0.195\textwidth]{
    \includegraphics[height=1in]{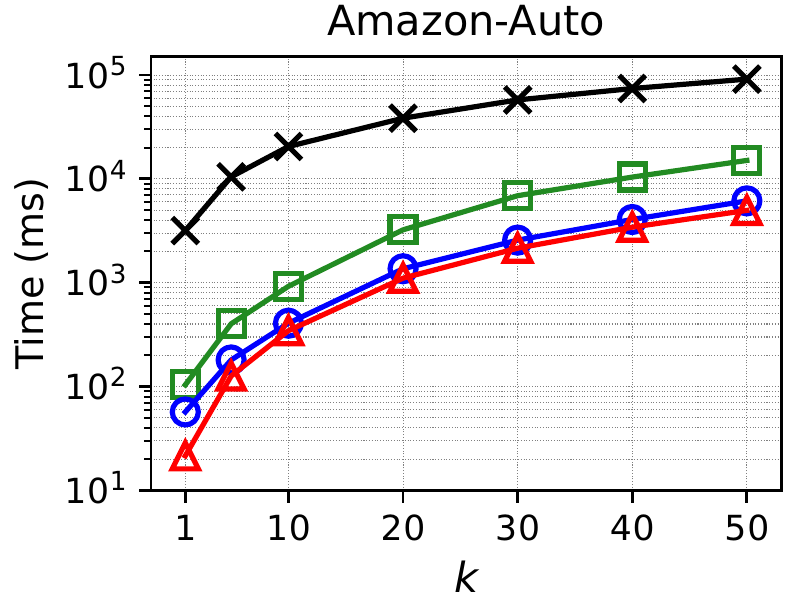}
  }
  \subcaptionbox{}[0.195\textwidth]{
    \includegraphics[height=1in]{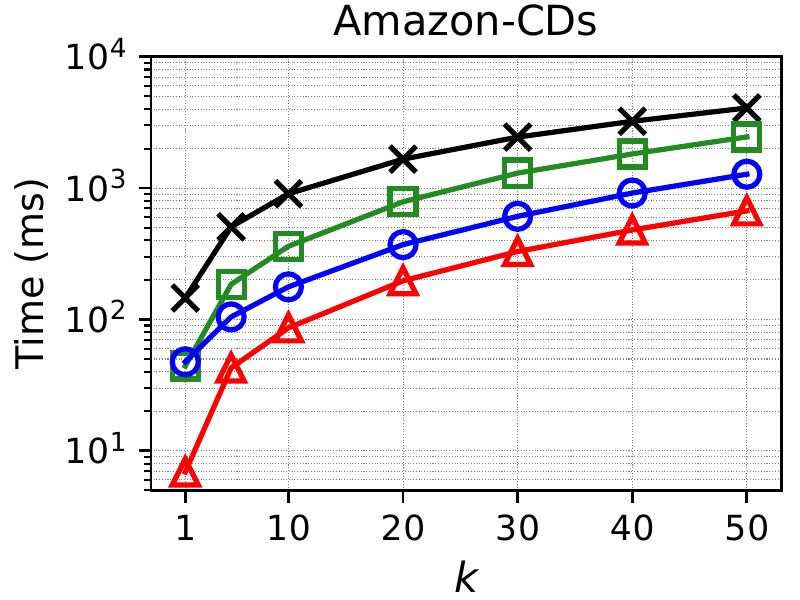}
  }
  \subcaptionbox{}[0.195\textwidth]{
    \includegraphics[height=1in]{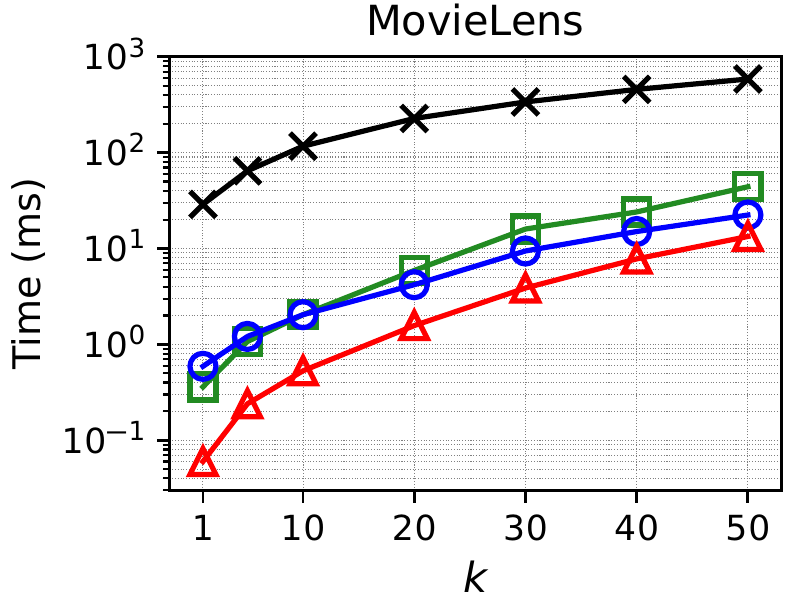}
  }
  \subcaptionbox{}[0.195\textwidth]{
    \includegraphics[height=1in]{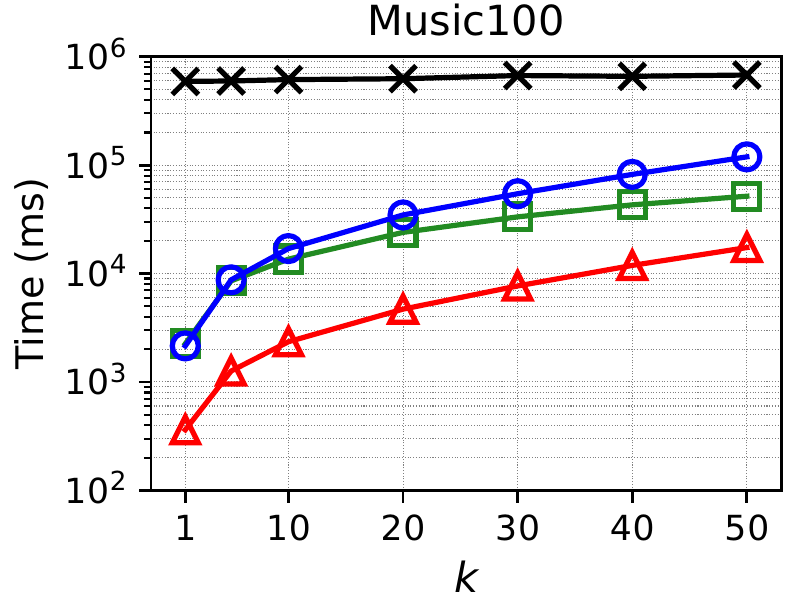}
  }
  \subcaptionbox{}[0.195\textwidth]{
    \includegraphics[height=1in]{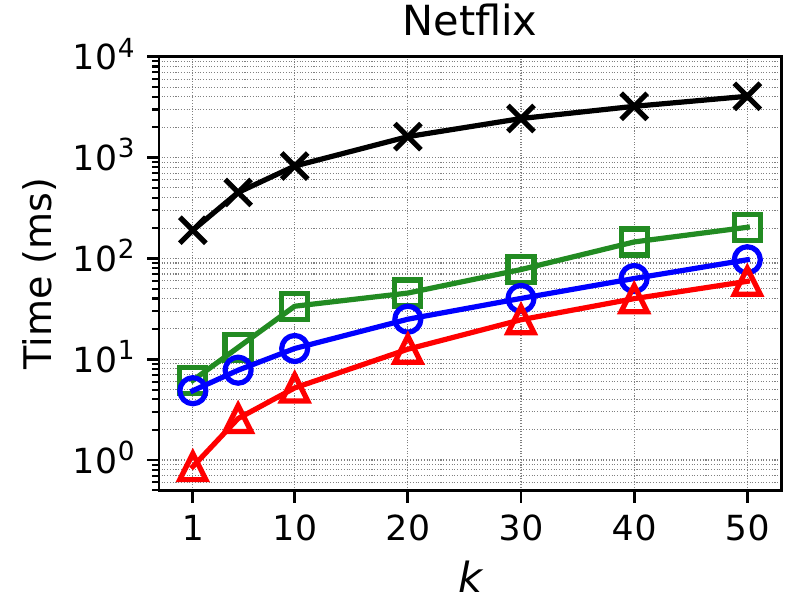}
  }
  \\ \vspace{-1em}
  \subcaptionbox{}[0.195\textwidth]{
    \includegraphics[height=1in]{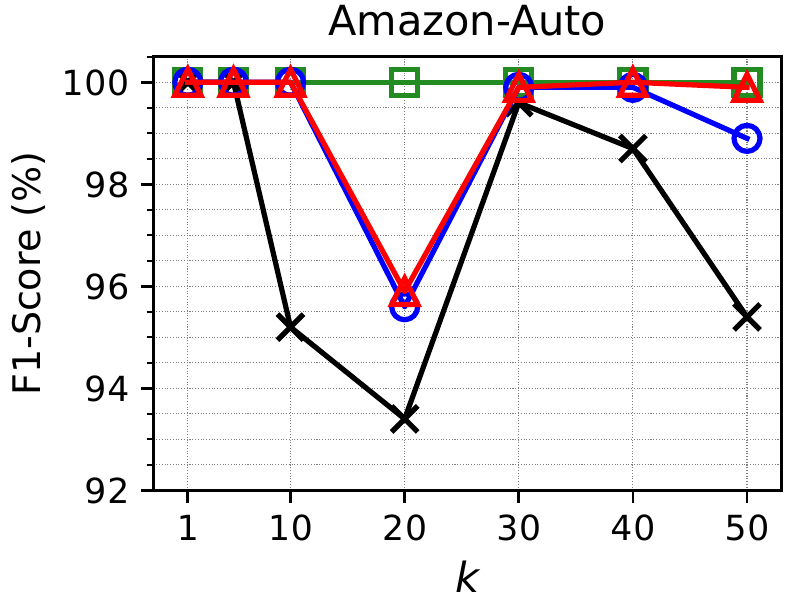}
  }
  \subcaptionbox{}[0.195\textwidth]{
    \includegraphics[height=1in]{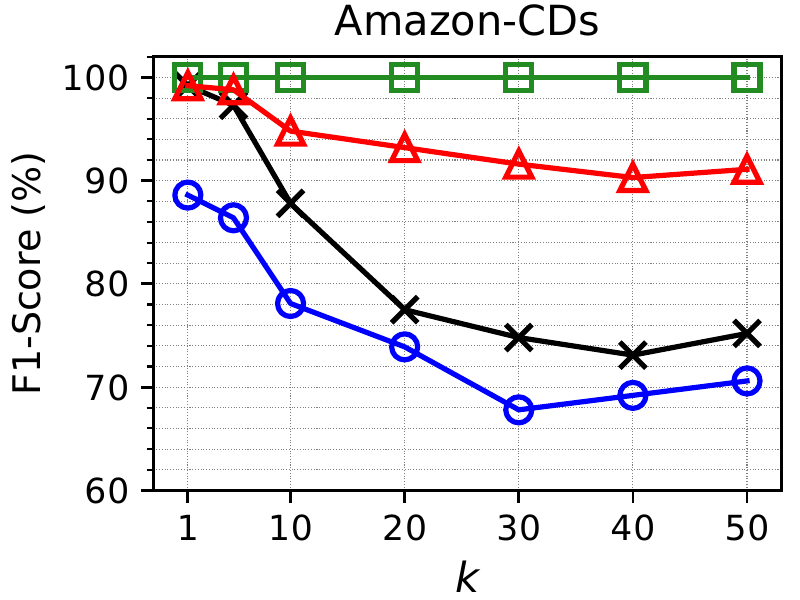}
  }
  \subcaptionbox{}[0.195\textwidth]{
    \includegraphics[height=1in]{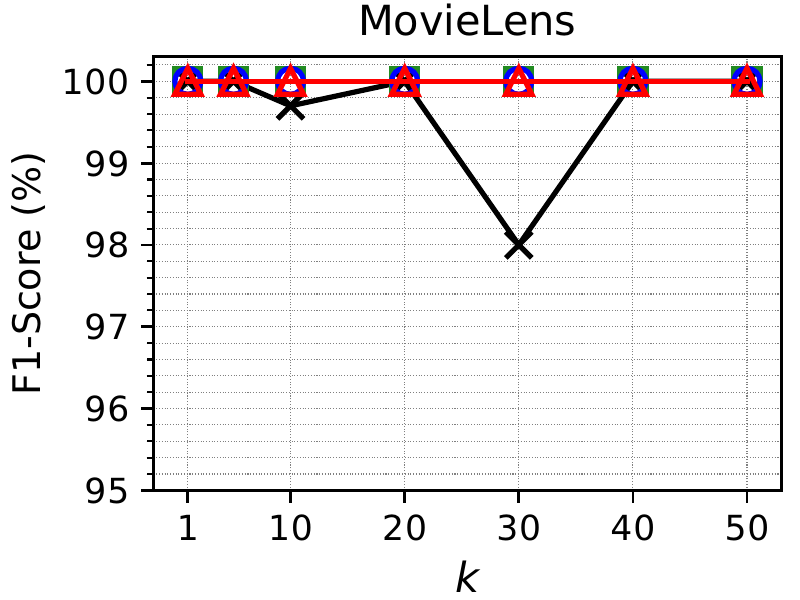}
  }
  \subcaptionbox{}[0.195\textwidth]{
    \includegraphics[height=1in]{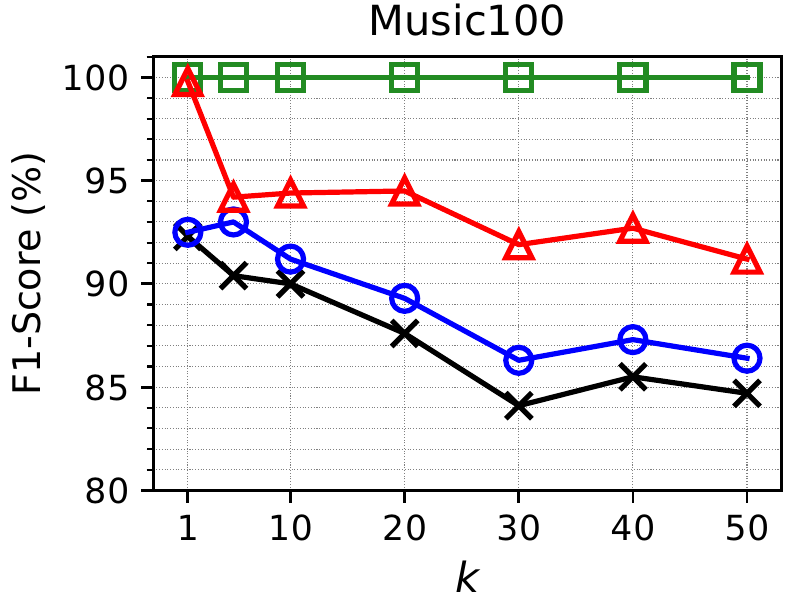}
  }
  \subcaptionbox{}[0.195\textwidth]{
    \includegraphics[height=1in]{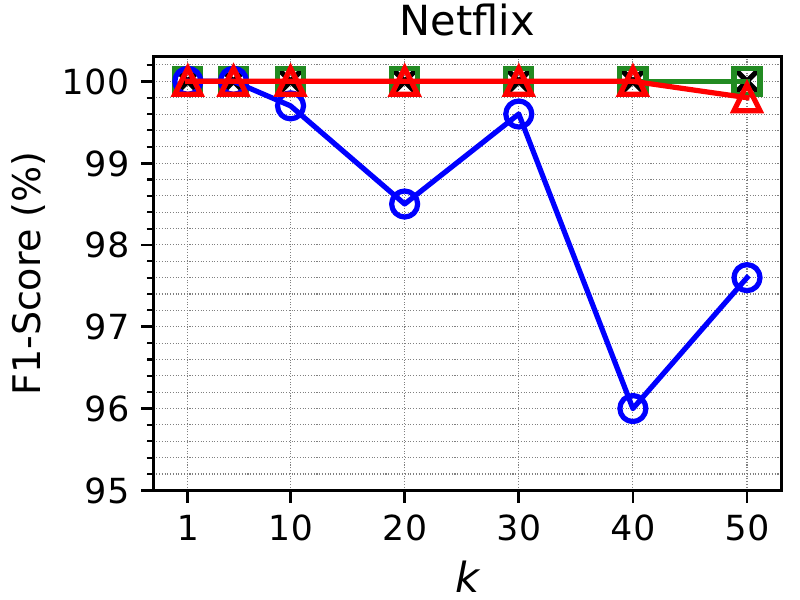}
  }
  \\ \vspace{-1em}
  \subcaptionbox{}[0.195\textwidth]{
    \includegraphics[height=1in]{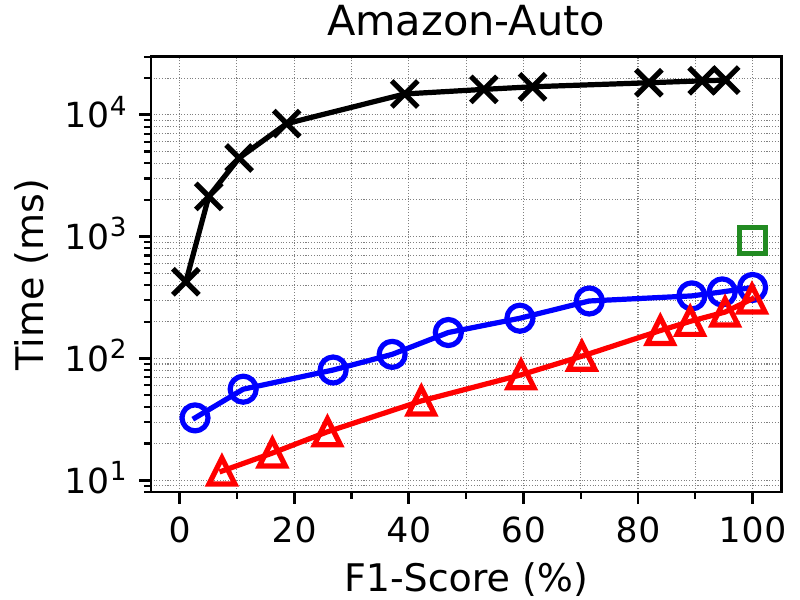}
  }
  \subcaptionbox{}[0.195\textwidth]{
    \includegraphics[height=1in]{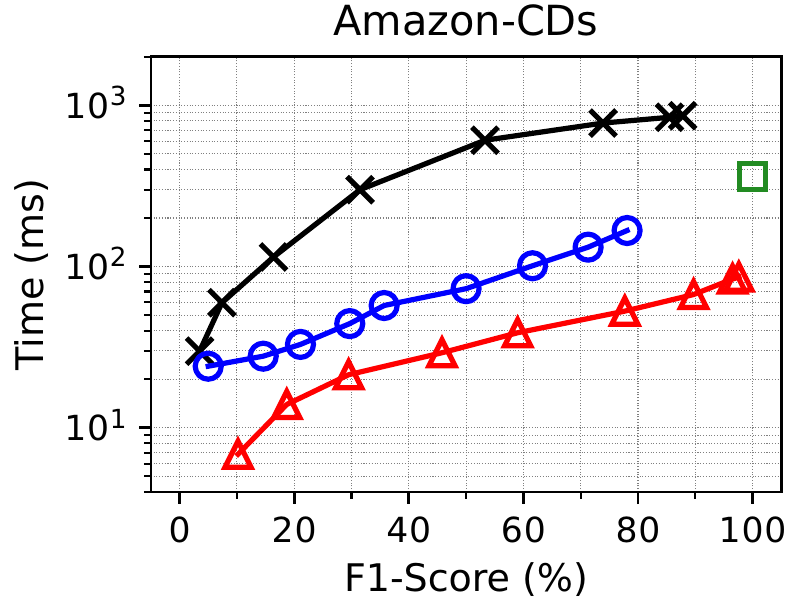}
  }
  \subcaptionbox{}[0.195\textwidth]{
    \includegraphics[height=1in]{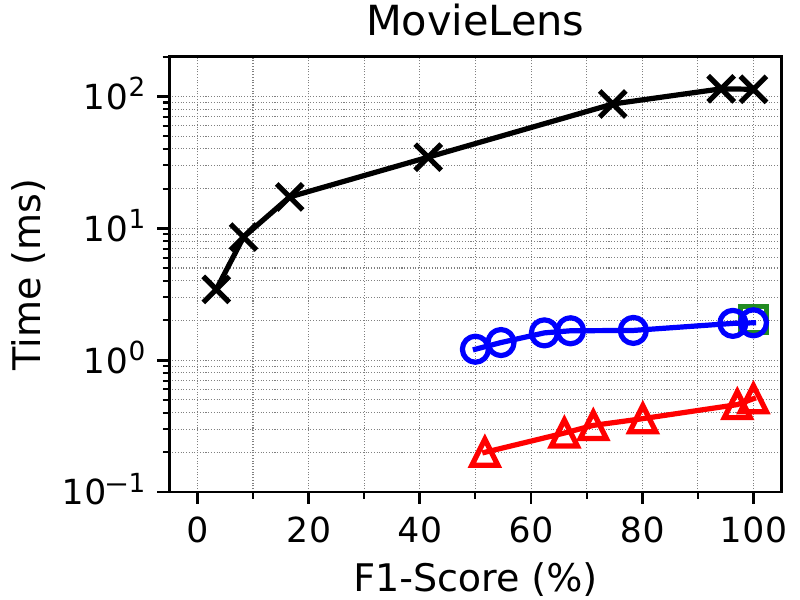}
  }
  \subcaptionbox{}[0.195\textwidth]{
    \includegraphics[height=1in]{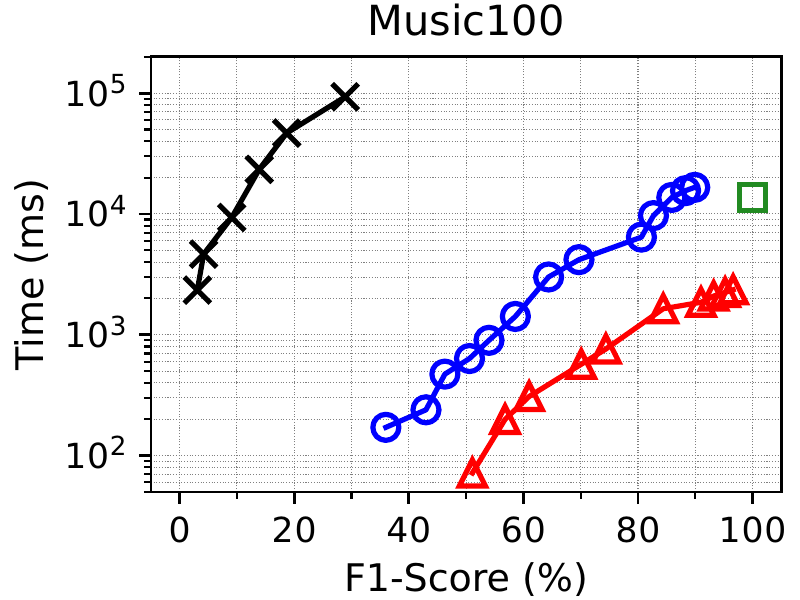}
  }
  \subcaptionbox{}[0.195\textwidth]{
    \includegraphics[height=1in]{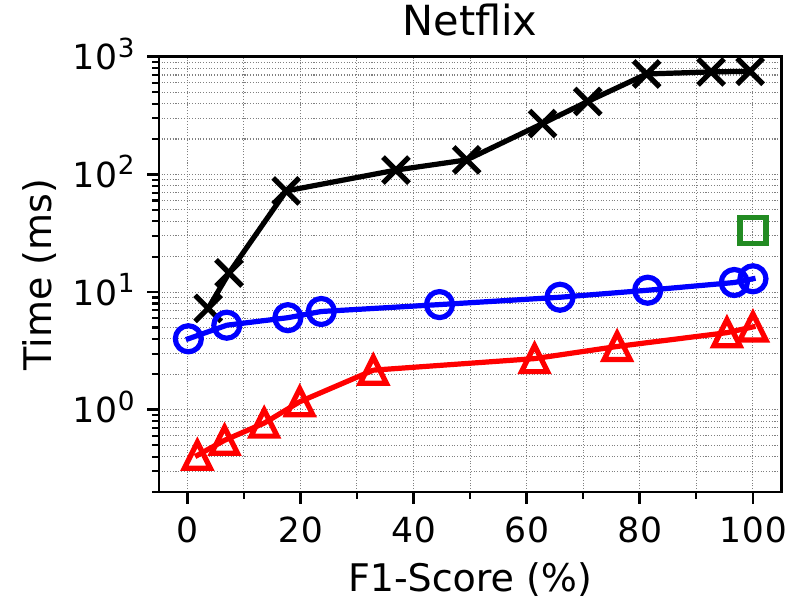}
  }
  \caption{Query performance of different algorithms.}
  \label{fig:exp:query}
  \vspace{-1.0em}
\end{figure*}

\paragraph{Setup.}
We evaluate the performance of SAH for R$k$MIPS through extensive experiments. 
We compare SAH with one state-of-the-art $k$MIPS method H2-ALSH\footnote{\url{https://github.com/HuangQiang/H2_ALSH}} and the only known R$k$MIPS method Simpfer.\footnote{\url{https://github.com/amgt-d1/Simpfer}}
To provide a more systematic comparison, we integrate the R$k$MIPS optimizations of Simpfer into H2-ALSH as a new baseline called H2-Simpfer.
All methods are implemented in C++ and compiled by g++-8 using -O3 optimization.
We conduct all experiments on a server with an Intel\textsuperscript{\textregistered} Xeon\textsuperscript{\textregistered} Platinum 8170 CPU @ 2.10GHz and 512 GB memory, running on CentOS 7.4. Each method is run on a single thread.

In the experiments, we use five real-world recommendation datasets, i.e.,
Amazon-Auto,\footnote{\url{https://nijianmo.github.io/amazon/index.html}} 
Amazon-CDs,\footnote{\url{http://jmcauley.ucsd.edu/data/amazon/index_2014.html}}
MovieLens,\footnote{\url{https://grouplens.org/datasets/movielens/}}
Music100~\cite{DBLP:conf/nips/MorozovB18}, and
Netflix~\cite{bennett2007netflix}.
The numbers of item and user vectors $(n,m)$ in Amazon-Auto, Amazon-CDs, MovieLens, Music100, and Netflix are ($925387$, $3873247$), ($64443$, $75258$), ($10681$, $71567$), ($1000000$, $1000000$), and ($17770$, $480189$), respectively.
The dimensionality $d$ of each dataset is $100$.
We randomly select $100$ item vectors as queries for each dataset.
The detailed procedures of dataset and query generation are described in the appendix. 

We use the query time to evaluate search efficiency and the F1-score to assess search accuracy.
For SAH, we use 128 hash tables in SA-ALSH and set the leaf size $N_0 = 20$ in Cone-Tree.
We fix $b=0.5$ for SAH, H2-ALSH, and H2-Simpfer and set $k_{max}=50$ for SAH, Simpfer, and H2-Simpfer; for other parameters in Simpfer and H2-ALSH, we use their default values \cite{DBLP:conf/recsys/AmagataH21,DBLP:conf/kdd/HuangMFFT18}.
All results are averaged by repeating each experiment five times with different random seeds.

\paragraph{Query Performance.}
We evaluate the performance of all methods for R$k$MIPS with varying $k \in \{1, 5, 10, 20, 30, 40$, $50\}$.
The results are shown in the first two rows of Figure~\ref{fig:exp:query}.

From the first row of Figure \ref{fig:exp:query}, we observe that SAH is about 4$\sim$8$\times$ faster than Simpfer. This is because SAH leverages SA-ALSH to conduct the $k$MIPS on item vectors, each in $O(d n^\rho \log n)$ time, where $0 < \rho < 1$. In contrast, Simpfer retrieves the $k$MIPS results by performing a linear scan over all item vectors, which requires $O(nd)$ time.
Compared with H2-ALSH, the advantage of SAH is more apparent: it runs nearly or over two orders of magnitude faster than H2-ALSH.
In particular, on the Music100 dataset, H2-ALSH is about three orders of magnitude slower than other methods, and it requires taking more than one day to complete all queries.
This observation justifies that leveraging the existing hashing schemes (e.g., H2-ALSH) to speed up the $k$MIPS for each user vector is not efficient for R$k$MIPS.
It also validates the effectiveness of our Cone-Tree blocking for pruning user vectors.
Besides, SAH always runs (up to 8$\times$) faster than H2-Simpfer. Since both methods utilize sublinear-time algorithms for $k$MIPS, this finding further confirms the effectiveness of our pruning strategies based on the cone structure.

Furthermore, we plot the curves of F1-score vs.~$k$ of all methods in the second row of Figure~\ref{fig:exp:query}. 
As Simpfer is an exact R$k$MIPS method, its F1-scores are always 100\%. 
The F1-scores of SAH across all datasets are over 90\% and consistently higher than those of H2-ALSH and H2-Simpfer. This advantage is attributed to the reductions in distortion errors coming from the shifting-invariant asymmetric transformation compared with the QNF transformation used in H2-ALSH and H2-Simpfer.

\begin{figure*}[ht]
  \captionsetup{skip=3pt}
  \captionsetup[sub]{labelformat=empty,skip=0pt,position=top,font={scriptsize}}
  \centering
  \includegraphics[height=0.12in]{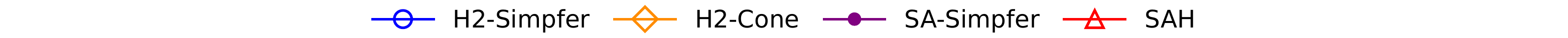}
  \\ \vspace{-1em}
  \subcaptionbox{}[0.195\textwidth]{
    \includegraphics[height=1in]{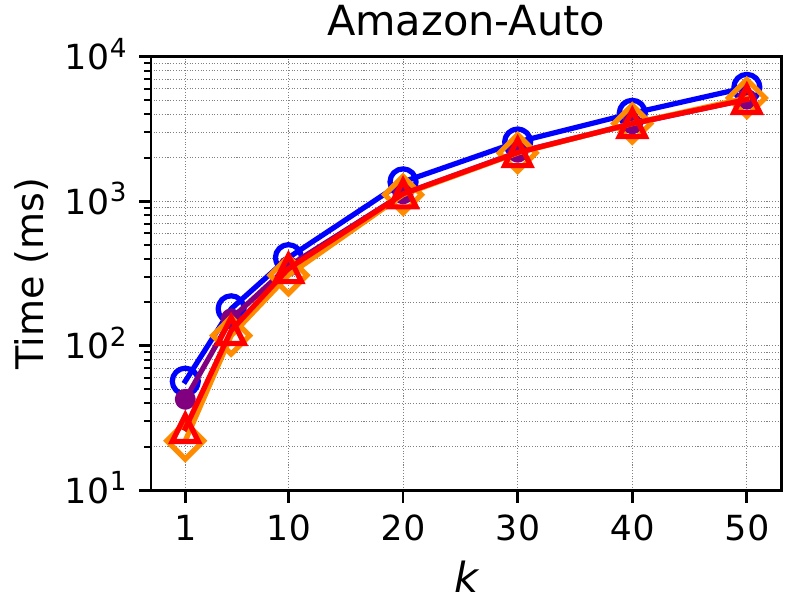}
  }
  \subcaptionbox{}[0.195\textwidth]{
    \includegraphics[height=1in]{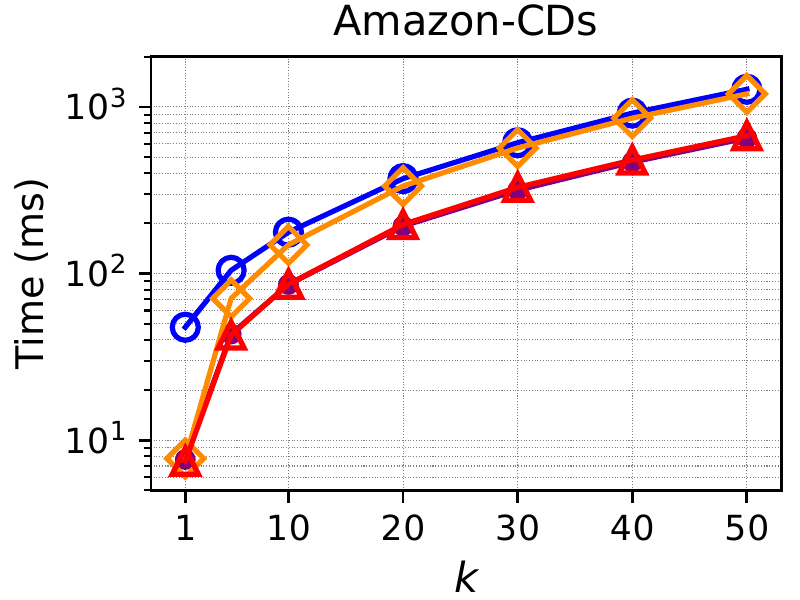}
  }
  \subcaptionbox{}[0.195\textwidth]{
    \includegraphics[height=1in]{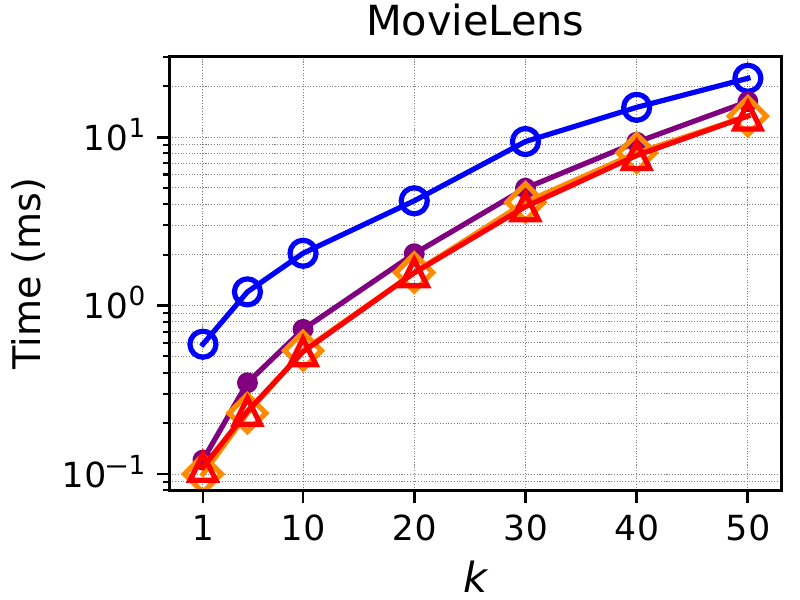}
  }
  \subcaptionbox{}[0.195\textwidth]{
    \includegraphics[height=1in]{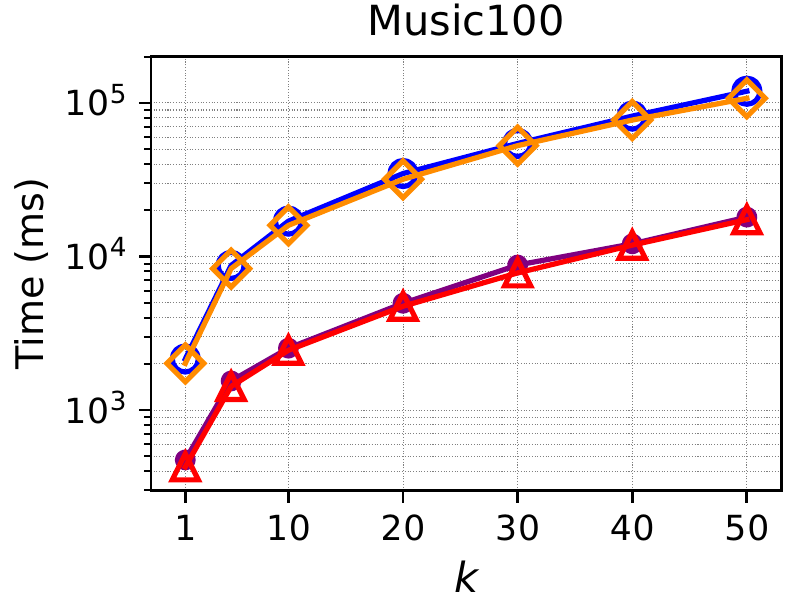}
  }
  \subcaptionbox{}[0.195\textwidth]{
    \includegraphics[height=1in]{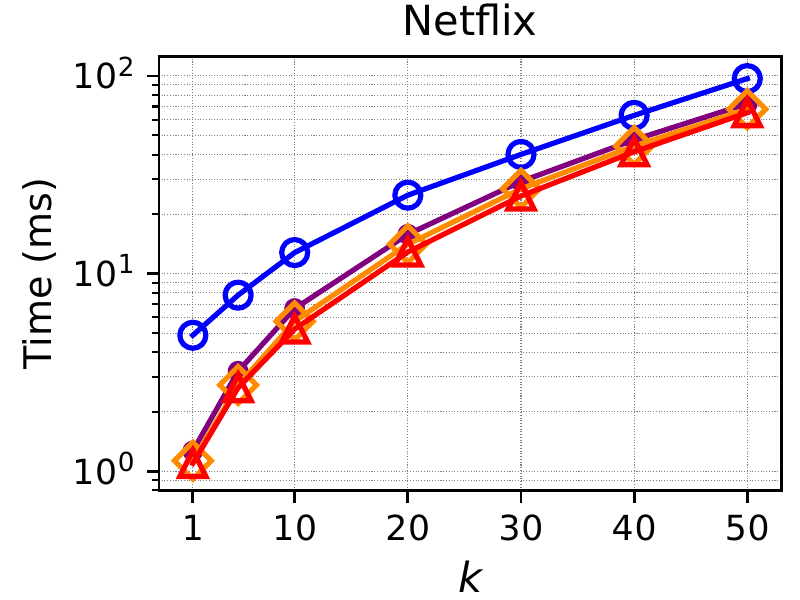}
  }
  \\ \vspace{-1em}
  \subcaptionbox{}[0.195\textwidth]{
    \includegraphics[height=1in]{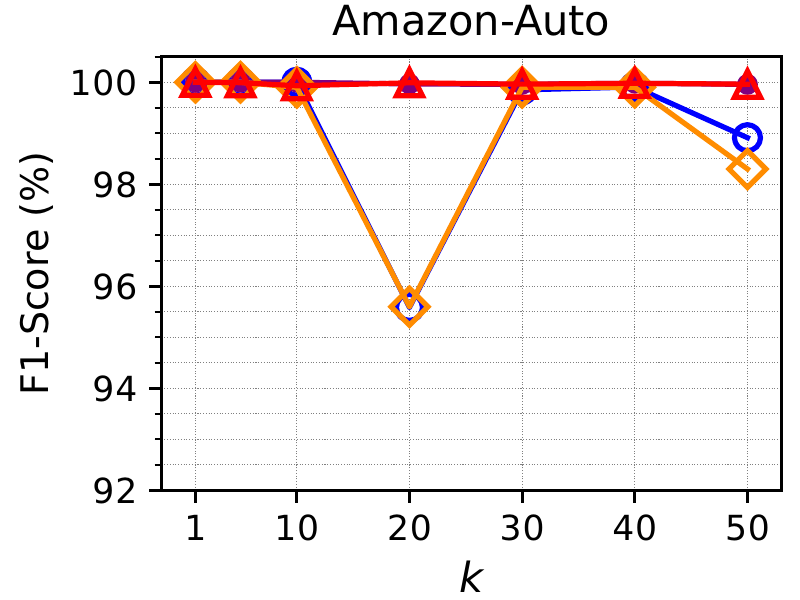}
  }
  \subcaptionbox{}[0.195\textwidth]{
    \includegraphics[height=1in]{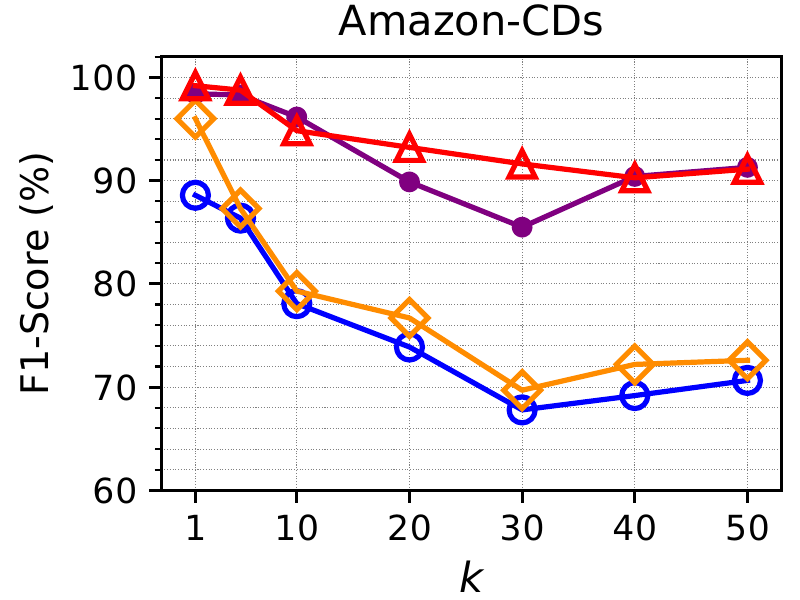}
  }
  \subcaptionbox{}[0.195\textwidth]{
    \includegraphics[height=1in]{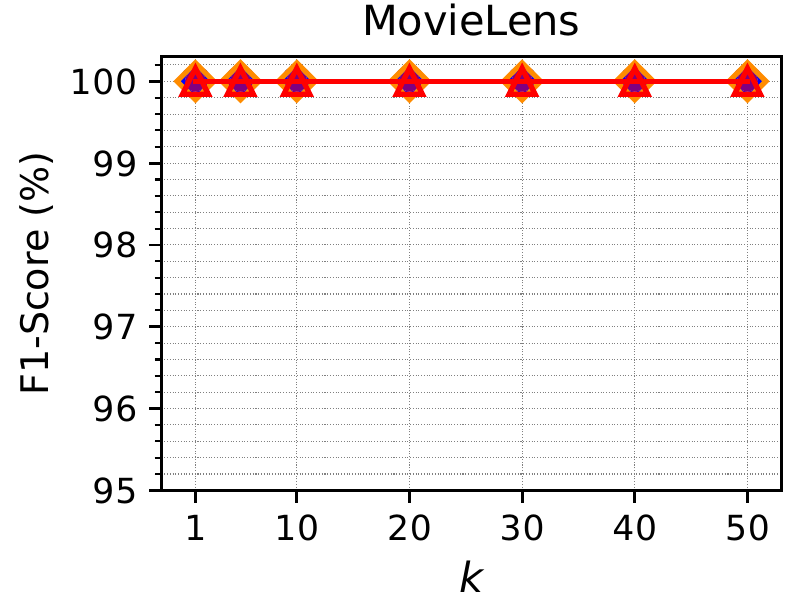}
  }
  \subcaptionbox{}[0.195\textwidth]{
    \includegraphics[height=1in]{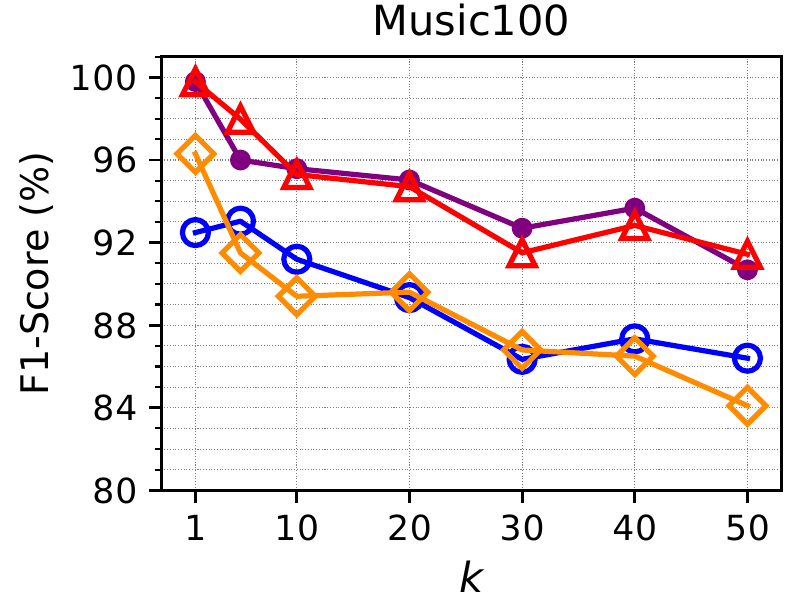}
  }
  \subcaptionbox{}[0.195\textwidth]{
    \includegraphics[height=1in]{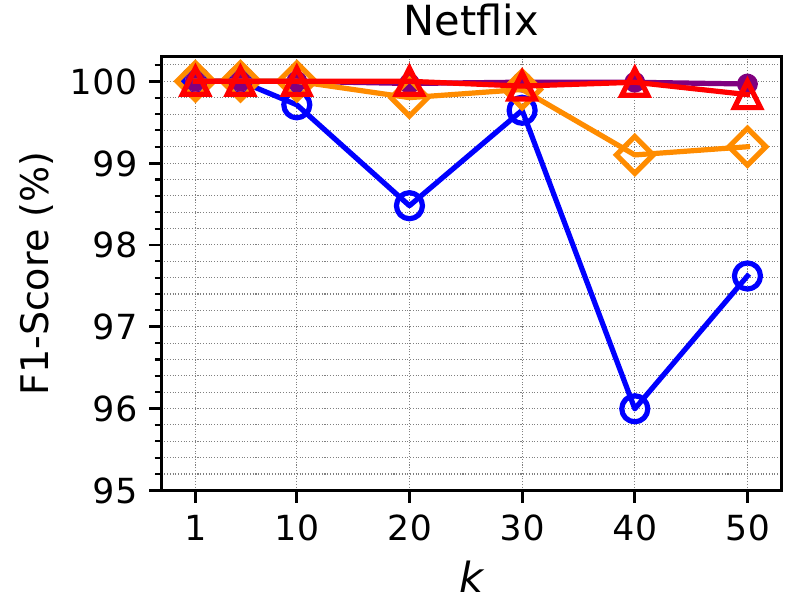}
  }
  \caption{Ablation Study for SAH with varying $k \in \{1, 5, 10, 20, 30, 40, 50\}$.}
  \label{fig:exp:ablation}
  \vspace{-1.0em}
\end{figure*}

Finally, we illustrate the query time of each algorithm as a function of F1-score when $k=10$ to compare their trade-offs between time efficiency and query accuracy in the third row of Figure~\ref{fig:exp:query}.
For each LSH-based algorithm, we present its query time and F1-score by varying the number of probed buckets until either the F1-score reaches 100\%, or the total number of probed buckets exceeds 50\%.
Since Simpfer is an exact algorithm without hash-based partitioning, we only plot its query time @100\% F1-score.
The results are consistent with the ones for varying $k$ in the first two rows of Figure~\ref{fig:exp:query}.
We observe that (1) SAH uniformly achieves the best trade-off between efficiency and accuracy in all cases; (2) the query time of SAH is still much lower than that of Simpfer when its F1-score approaches 100\%.
These results confirm the superior effectiveness and efficiency of SAH for R$k$MIPS in a more detailed manner.

\begin{table}[t]
\captionsetup{skip=3pt}
\centering
\caption{Indexing time (in seconds) of each algorithm.}
\label{tbl:index:time}
\resizebox{\columnwidth}{!}{%
\begin{tabular}{cccccc}
\toprule
\textbf{Dataset} & Simpfer & H2-ALSH & H2-Simpfer & SAH \\
\midrule
Amazon-Auto & 329.8 & \textbf{19.7} & 294.1 & 428.0 \\
Amazon-CDs & 9.4 & \textbf{1.3} & 9.7 & 11.5 \\
MovieLens & 2.6 & \textbf{0.1} & 1.9 & 2.7 \\
Music100 & 32.8 & \textbf{9.3} & 32.2 & 46.6 \\
Netflix & 18.5 & \textbf{0.2} & 13.3 & 20.8 \\
\bottomrule
\end{tabular}
}
\vspace{-0.5em}
\end{table}

\paragraph{Indexing Time.}
We present the indexing time of each algorithm in Table~\ref{tbl:index:time}.
H2-ALSH always takes the least time for index construction because it only builds an index on items.
Nevertheless, as a trade-off, it cannot prune any user vector in the R$k$MIPS processing. Thus, its query efficiency is not comparable to other algorithms for solving R$k$MIPS.
The indexing time of SAH is slightly (1.05$\sim$1.43$\times$) larger than that of Simpfer, primarily because of the additional cost of building the Cone-Tree.
Nevertheless, the indexing phase of SAH can be completed within 7.2 minutes, even on Amazon-Auto with nearly 1M items and 3.8M users. This finding indicates that the index construction of SAH is scalable to large datasets.

\paragraph{Ablation Study.}
We conduct an ablation study for SAH. To validate the effectiveness of SA-ALSH and the Cone-Tree blocking strategy separately, we first remove the Cone-Tree blocking strategy from SAH and integrate the Simpfer optimizations into SA-ALSH as a new baseline called SA-Simpfer. Then, we replace SA-ALSH with H2-ALSH while retaining the Cone-Tree blocking strategy as another baseline called H2-Cone. We show the results of H2-Simpfer, H2-Cone, SA-Simpfer, and SAH in Figure~\ref{fig:exp:ablation}.

From the first row of Figure \ref{fig:exp:ablation}, we discover that H2-Simpfer takes longer query time than SA-Simpfer in almost all cases. This observation validates that SA-ALSH is more efficient than H2-ALSH for performing $k$MIPS, which would be because SA-ALSH incurs less distortion error than H2-ALSH, so that it finds the $k$MIPS as early as possible and triggers the upper bound $\mu_j = M_j \norm{\bm{u}}$ of the user vector $\bm{u}$ for early pruning.
Moreover, we find that the query time of H2-Cone and SAH is uniformly shorter than H2-Simpfer and SA-Simpfer, respectively. This finding confirms the effectiveness of the Cone-Tree blocking compared with the norm-based blocking in Simpfer.
For some datasets such as Amazon-Auto and Music100, the advantage of the Cone-Tree blocking is not very apparent because the effectiveness of upper bounds in Lemmas~\ref{lemma:node_upper_bound} and~\ref{lemma:point_upper_bound} to prune unnecessary inner product evaluations relies on the angle distribution of user vectors and might be less useful when the angles between most pairs of user vectors are close to $\pi/2$.

From the second row of Figure \ref{fig:exp:ablation}, we discover that the F1-scores of H2-Simpfer and H2-Cone are worse than those of SA-Simpfer and SAH, and their results are also less stable. This discovery empirically justifies the effectiveness of SAT in reducing the distortion error so that SA-Simpfer and SAH, where SA-ALSH is used for $k$MIPS, have higher F1-scores than H2-Simpfer and H2-Cone.
These results are consistent with Figure~\ref{fig:exp:query}.
Finally, H2-Cone and SA-Simpfer are inferior to SAH in almost all cases, which validates that the integration of SA-ALSH and Cone-Tree based blocking further improves the performance upon using them individually with existing blocking methods or ALSH schemes.

In addition, we also study how different parameters affect the performance of SAH and validate the effectiveness of SA-ALSH for $k$MIPS. The results and analyses are left to the appendix. 

\section{Conclusion}
\label{sec:conclusion}
In this paper, we studied a new yet difficult problem called R$k$MIPS on high-dimensional data.
We proposed the first subquadratic-time algorithm SAH to tackle the R$k$MIPS efficiently and effectively in two folds.
First, we developed a novel sublinear-time hashing scheme SA-ALSH to accelerate the $k$MIPS on item vectors. With the shifting-invariant asymmetric transformation, the distortion errors were reduced significantly.
Second, we devised a new Cone-Tree blocking strategy that effectively pruned user vectors (in a batch).
Extensive experiments on five real-world datasets confirmed the superior performance of SAH in terms of search accuracy and efficiency.
Our work will likely contribute to opening up a new research direction and providing a practical solution to this challenging problem.

In future work, since the SAH algorithm achieves a theoretical guarantee for solving R$k$MIPS only when $k=1$, it would be interesting to design a subquadratic-time algorithm for approximate R$k$MIPS with any $k>1$.

\section*{Acknowledgments}
We thank Yangyang Guo for helping us with dataset generation.
This research is supported by the National Research Foundation, Singapore under its Strategic Capability Research Centres Funding Initiative and the National Natural Science Foundation of China under grant No.~62202169.
Any opinions, findings, and conclusions, or recommendations expressed in this material are those of the author(s) and do not reflect the views of the National Research Foundation, Singapore.

\bibliography{aaai23}

\appendix

\section{Additional Experiments}
\label{sec:appendix:expt}

\begin{figure*}[t]
  \captionsetup{skip=3pt}
  \captionsetup[sub]{labelformat=empty,skip=0pt,position=top,font={scriptsize}}
  \centering
  \includegraphics[height=0.1in]{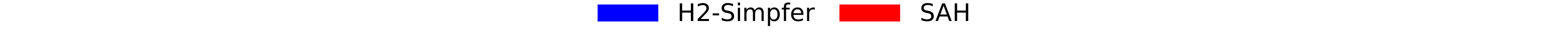}
  \\\vspace{-1em}
  \subcaptionbox{}[0.195\textwidth]{
    \includegraphics[height=1in]{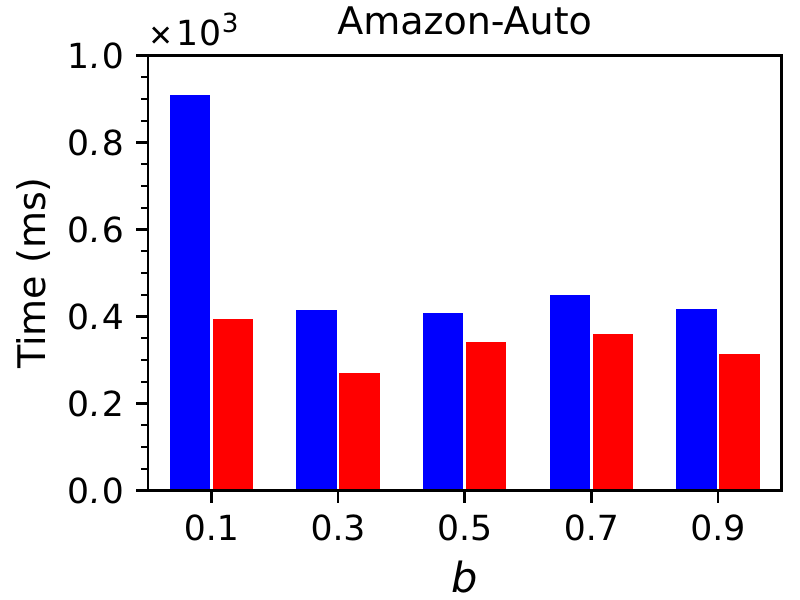}
  }
  \subcaptionbox{}[0.195\textwidth]{
    \includegraphics[height=1in]{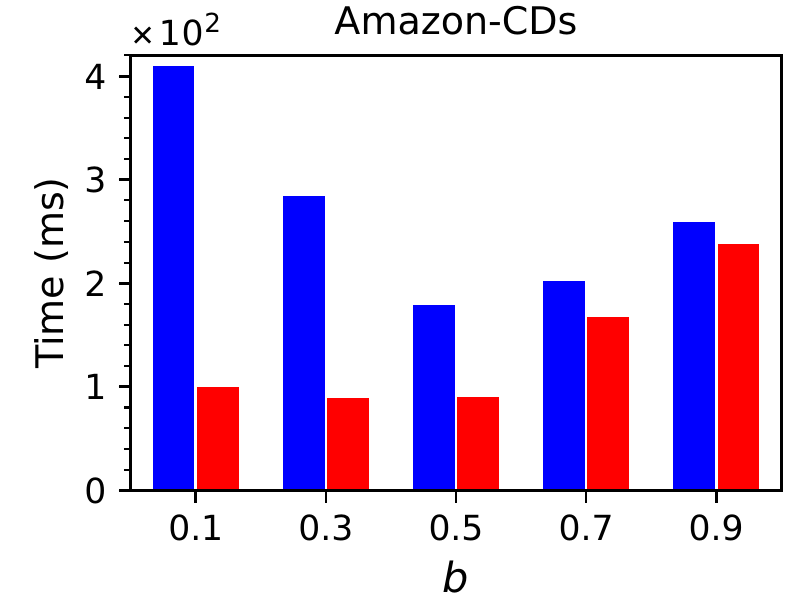}
  }
  \subcaptionbox{}[0.195\textwidth]{
    \includegraphics[height=1in]{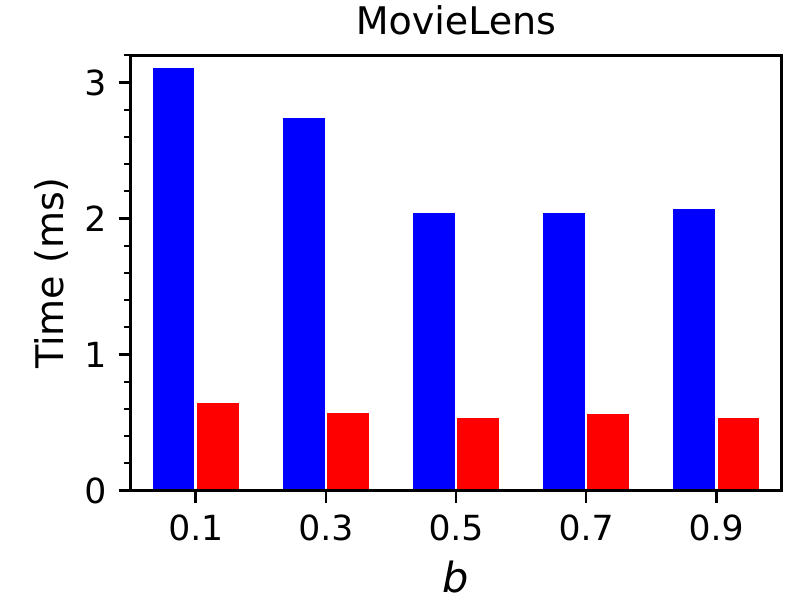}
  }
  \subcaptionbox{}[0.195\textwidth]{
    \includegraphics[height=1in]{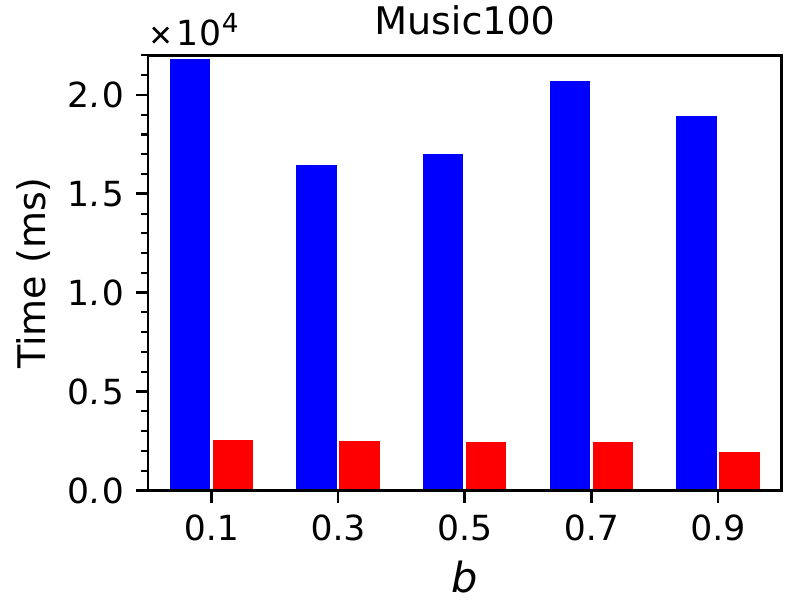}
  }
  \subcaptionbox{}[0.195\textwidth]{
    \includegraphics[height=1in]{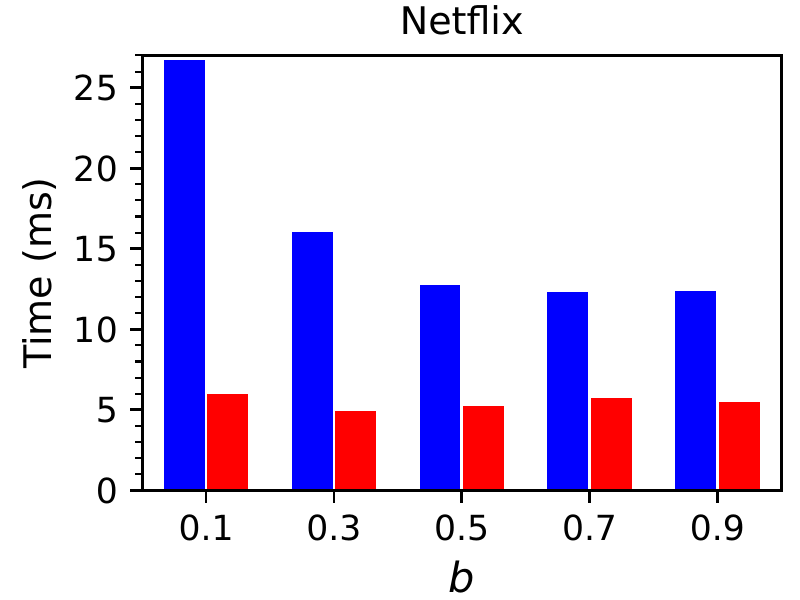}
  }
  \\\vspace{-1em}
  \subcaptionbox{}[0.195\textwidth]{
    \includegraphics[height=1in]{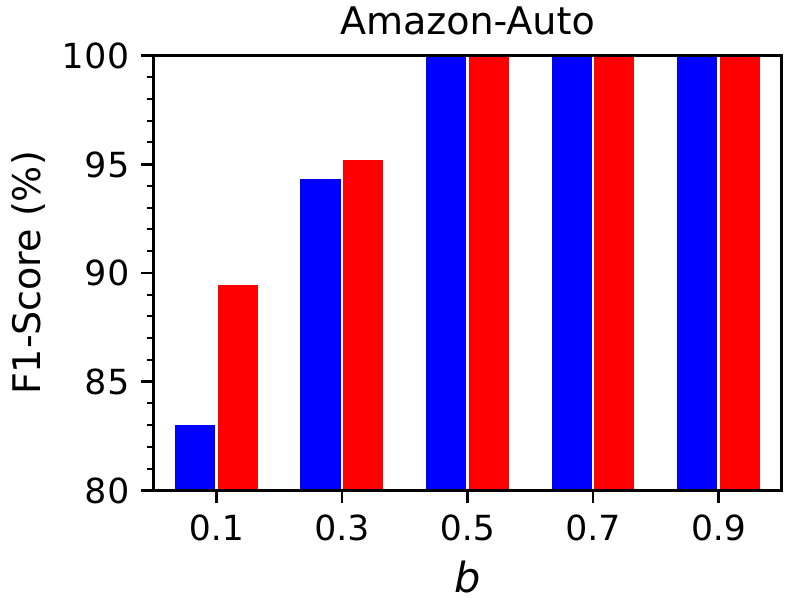}
  }
  \subcaptionbox{}[0.195\textwidth]{
    \includegraphics[height=1in]{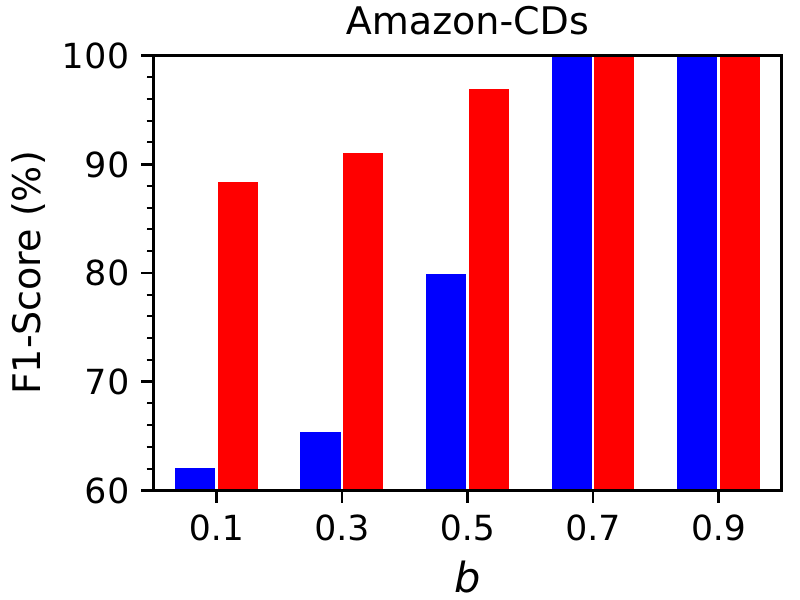}
  }
  \subcaptionbox{}[0.195\textwidth]{
    \includegraphics[height=1in]{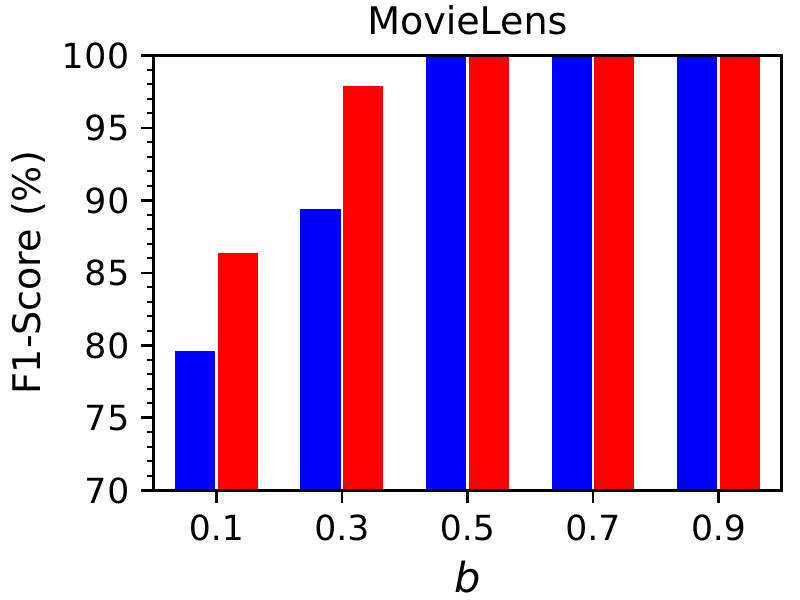}
  }
  \subcaptionbox{}[0.195\textwidth]{
    \includegraphics[height=1in]{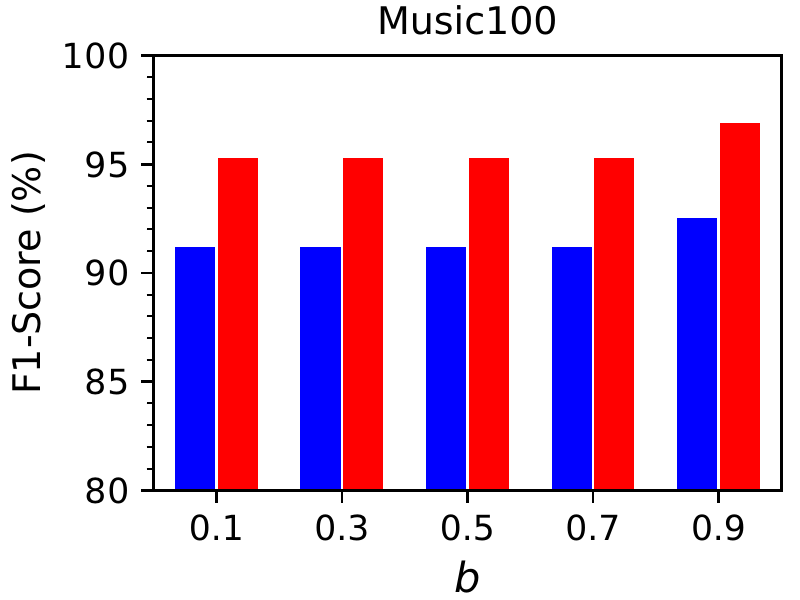}
  }
  \subcaptionbox{}[0.195\textwidth]{
    \includegraphics[height=1in]{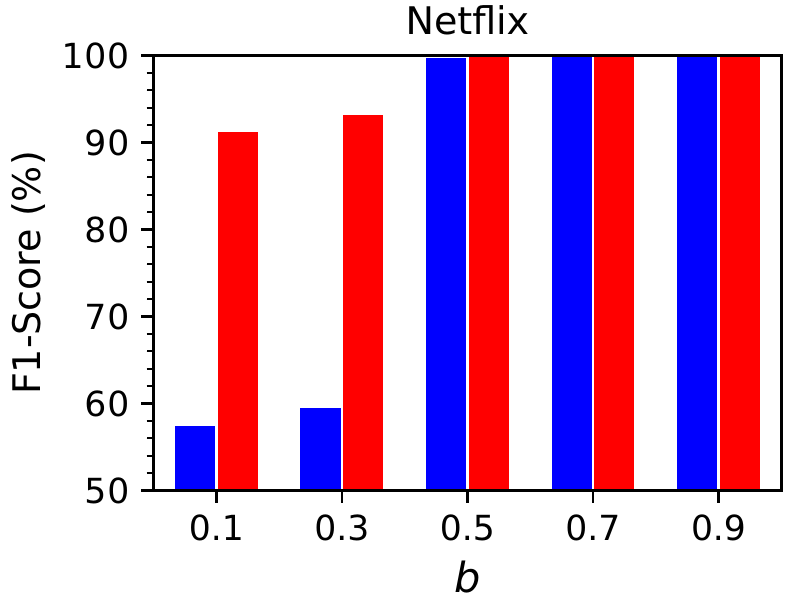}
  }
  \caption{Comparison between SAH and H2-Simpfer with varying $b \in \{0.1, 0.3, 0.5, 0.7, 0.9\}$.}
  \label{fig:exp:b}
  \vspace{-1.0em}
\end{figure*}

\begin{figure*}[ht]
  \captionsetup{skip=3pt}
  \captionsetup[sub]{labelformat=empty,skip=0pt,position=top,font={scriptsize}}
  \centering
  \subcaptionbox{}[0.195\textwidth]{
    \includegraphics[height=1in]{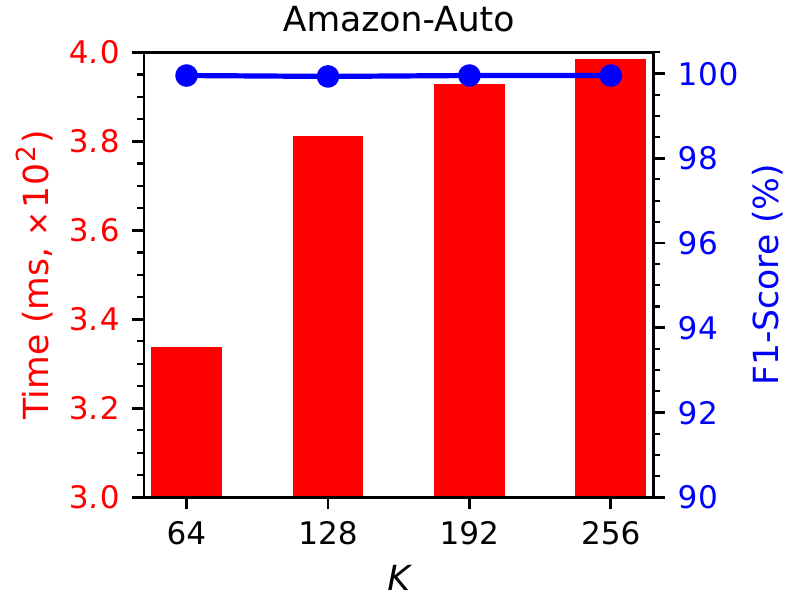}
  }
  \subcaptionbox{}[0.195\textwidth]{
    \includegraphics[height=1in]{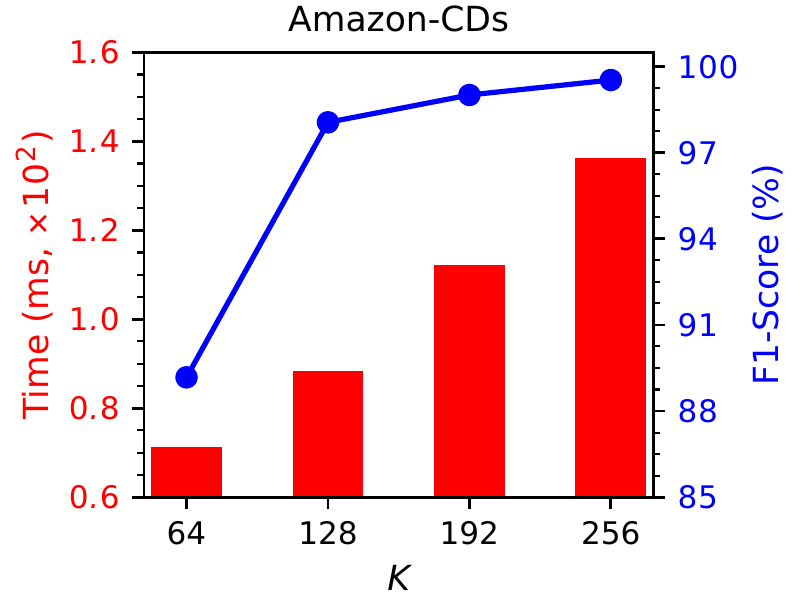}
  }
  \subcaptionbox{}[0.195\textwidth]{
    \includegraphics[height=1in]{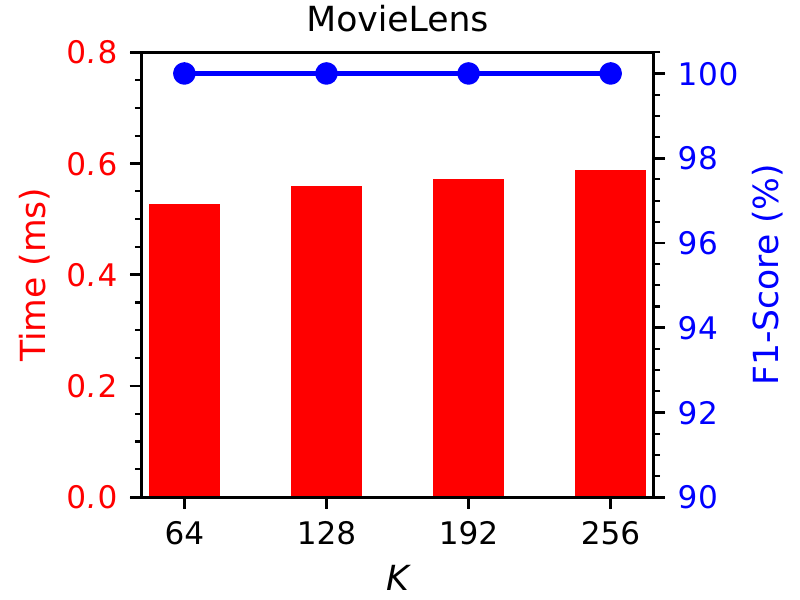}
  }
  \subcaptionbox{}[0.195\textwidth]{
    \includegraphics[height=1in]{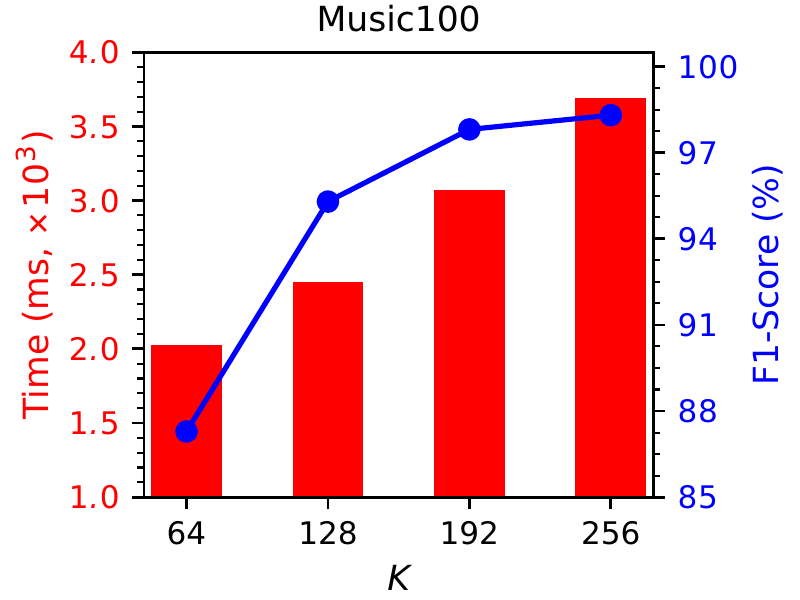}
  }
  \subcaptionbox{}[0.195\textwidth]{
    \includegraphics[height=1in]{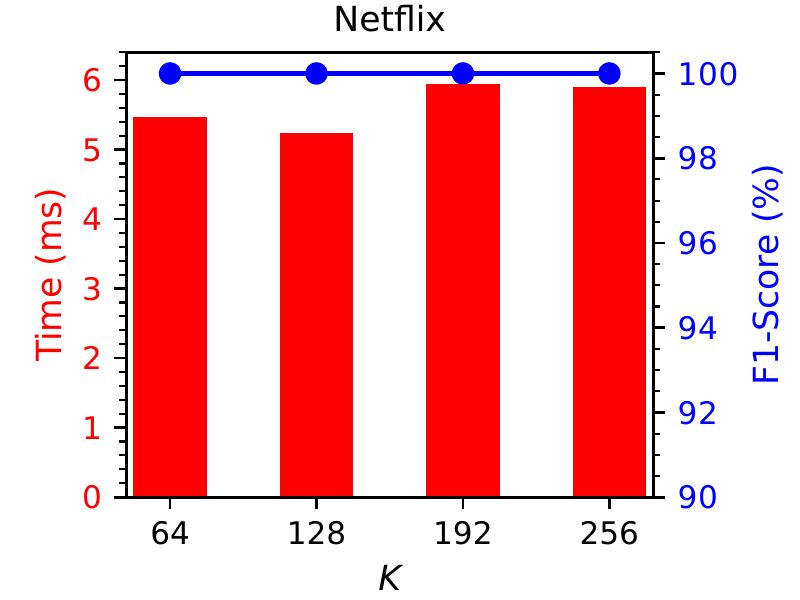}
  }
  \caption{Impact of the number of hash tables for SAH with varying $K \in \{64, 128, 192, 256\}$.}
  \label{fig:exp:K}
  \vspace{-0.5em}
\end{figure*}

\begin{figure*}[ht]
  \captionsetup{skip=3pt}
  \captionsetup[sub]{labelformat=empty,skip=0pt,position=top,font={scriptsize}}
  \centering
  \subcaptionbox{}[0.195\textwidth]{
    \includegraphics[height=1in]{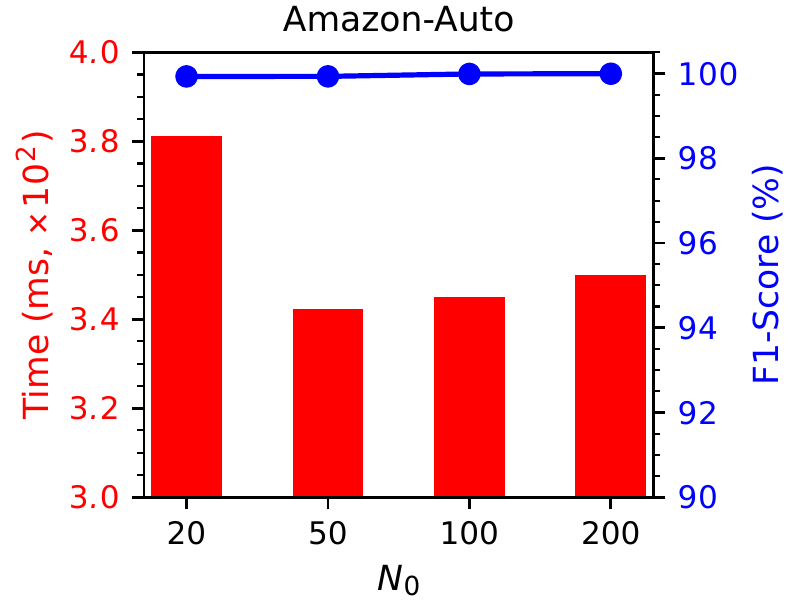}
  }
  \subcaptionbox{}[0.195\textwidth]{
    \includegraphics[height=1in]{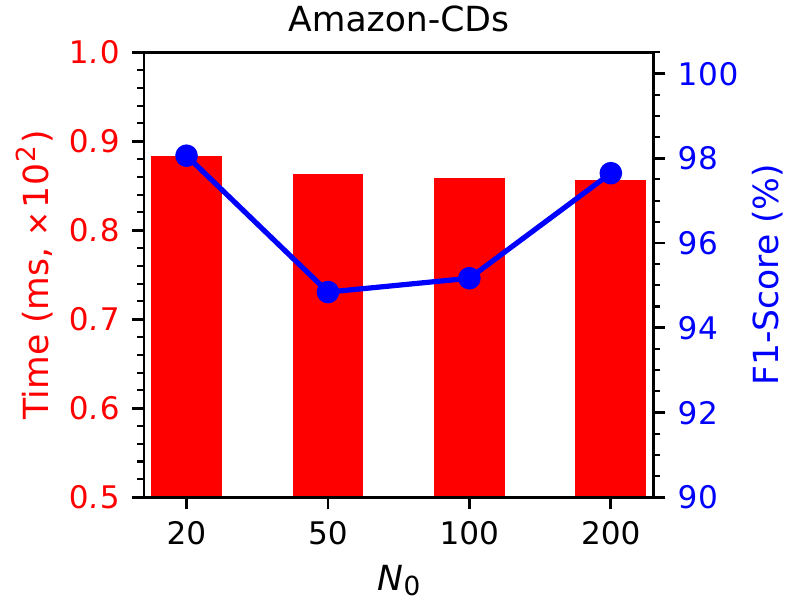}
  }
  \subcaptionbox{}[0.195\textwidth]{
    \includegraphics[height=1in]{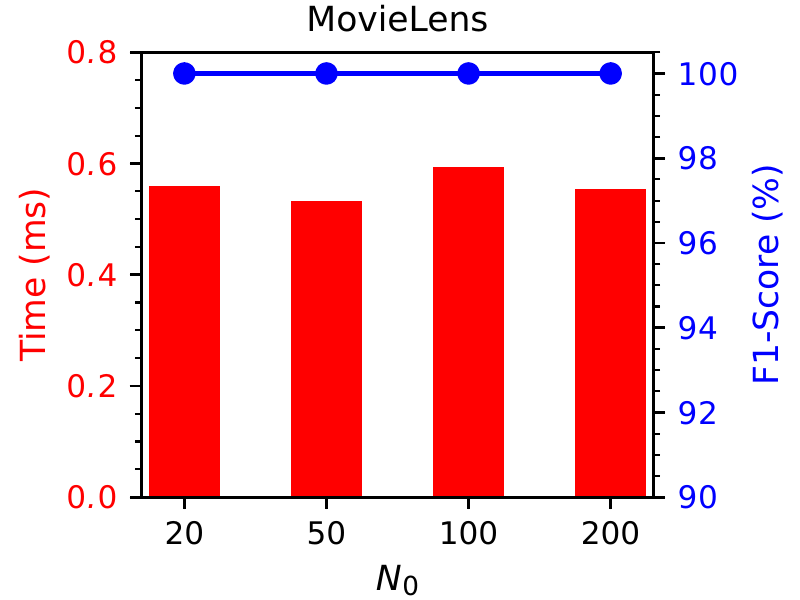}
  }
  \subcaptionbox{}[0.195\textwidth]{
    \includegraphics[height=1in]{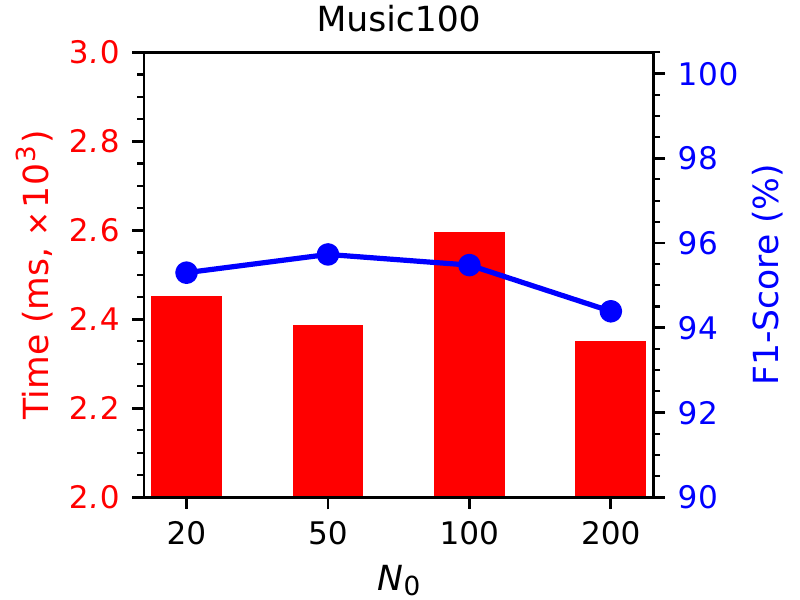}
  }
  \subcaptionbox{}[0.195\textwidth]{
    \includegraphics[height=1in]{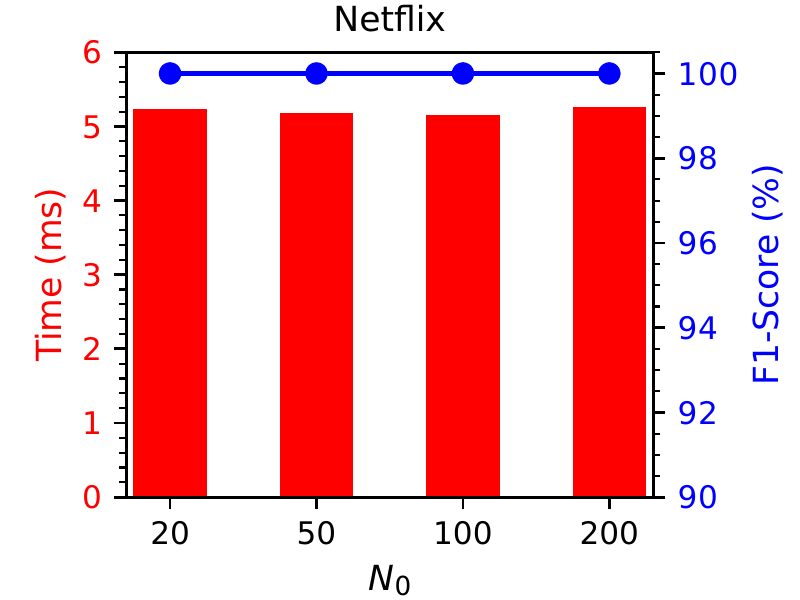}
  }
  \caption{Impact of the maximum leaf size for SAH with varying $N_0 \in \{20, 50, 100, 200\}$.}
  \label{fig:exp:N_0}
  \vspace{-0.5em}
\end{figure*}

\paragraph{Procedure for Dataset Generation.}
For the datasets Amazon-Auto, Amazon-CDs, MovieLens, and Netflix, we first retrieve their (sparse) user-item rating matrix $R$, where $R(i,j)$ is the rating of user $i$ for item $j$.
Then, we set up the latent dimensionality $d=100$ and apply the Non-negative Matrix Factorization (NMF)\footnote{\url{https://pytorch-nmf.readthedocs.io/en/stable/}} on $R$ to obtain their latent user and item matrices.
For the Music100 dataset, since the authors only provided a single dense matrix (not rating matrix) in~\cite{DBLP:conf/nips/MorozovB18}, we use it as both user and item matrices.
Finally, on each dataset, we randomly select $100$ item vectors from the item matrix and use them as the query set.
The dataset and queries are available at~\url{https://github.com/HuangQiang/SAH}.

\paragraph{Impact of the Interval Ratio $b$ for SAH and H2-Simpfer.}
We first study the impact of the interval ratio $b$ for SAH and H2-Simpfer. 
We fix $N_0=20$ and $K=128$ and vary $b \in \{0.1, 0.3, 0.5, 0.7, 0.9\}$. H2-ALSH is omitted since it is much worse than SAH and H2-Simpfer.
The results for R$k$MIPS with $k=10$ are displayed in Figure~\ref{fig:exp:b}. Similar trends can be observed for other values of $k$.

On the one hand, the query time of SAH is always less than that of H2-Simpfer.
On the other hand, the F1-scores of SAH are uniformly higher than those of H2-Simpfer. 
In overall, both SAH and H2-Simpfer with $b=0.5$ achieve the best trade-off between efficiency and accuracy among the five tested values. Thus, we set $b=0.5$ by default.
Moreover, the results are consistent with those in Figure 1 and confirm the effectiveness of the pruning strategies based on the cone structure as well as the shifting-invariant asymmetric transformation for reducing distortion errors.

\paragraph{Impact of the Number of Hash Tables $K$ for SAH.}
We study the impact of the number of hash tables $K$ in SA-ALSH for SAH.
We fix $b = 0.5$ and $N_0 = 20$ and vary $K \in \{64, 128, 192, 256\}$.
The results of SAH with $k=10$ are shown in Figure~\ref{fig:exp:K}.
Similar trends can be observed for other values of $k$.

From the bars in Figure~\ref{fig:exp:K}, we observe that the query time of SAH increases with $K$, and the increment is small from $K=128$ to $256$ in most datasets.
It is natural because, as $K$ increases, SAH takes more computational cost to obtain the hash values of queries and retrieve the candidates.
Moreover, from the curves of Figure~\ref{fig:exp:K}, the F1-scores of SAH also increase as $K$ increases, and the growth in the F1-scores of SAH is quite large from $K=64$ to $128$ on some datasets such as Amazon-CDs and Music100. It is also natural because, with more hash tables, SAH has a larger chance to find the correct results for $k$MIPS.
In summary, SAH with $K=128$ achieves the best trade-off between efficiency and accuracy among the four tested values, and $K=128$ is used by default.

\begin{figure*}[t]
  \captionsetup{skip=3pt}
  \captionsetup[sub]{labelformat=empty,skip=0pt,position=top,font={scriptsize}}
  \centering
  \includegraphics[height=0.12in]{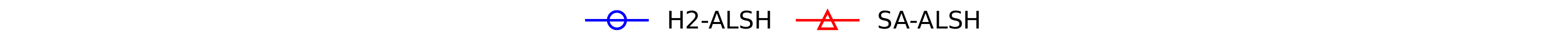}
  \\\vspace{-1em}
  \subcaptionbox{}[0.195\textwidth]{
    \includegraphics[height=1in]{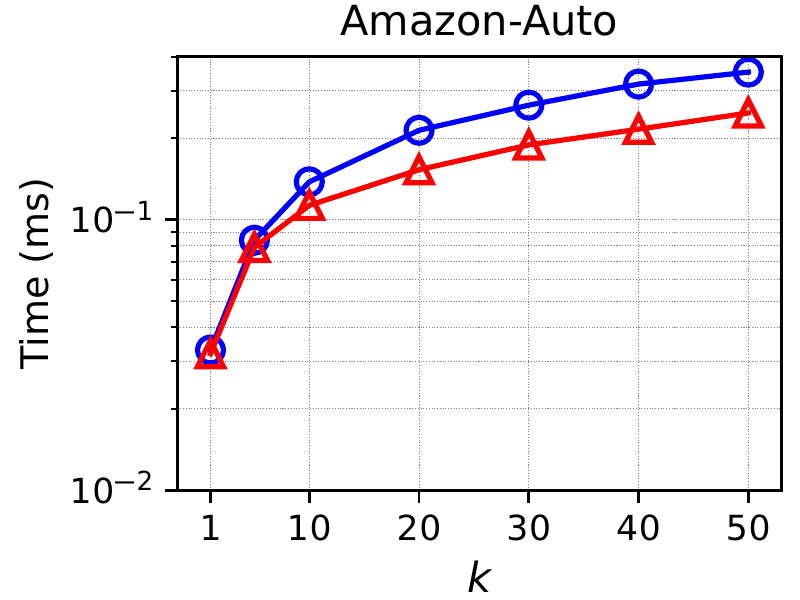}
  }
  \subcaptionbox{}[0.195\textwidth]{
    \includegraphics[height=1in]{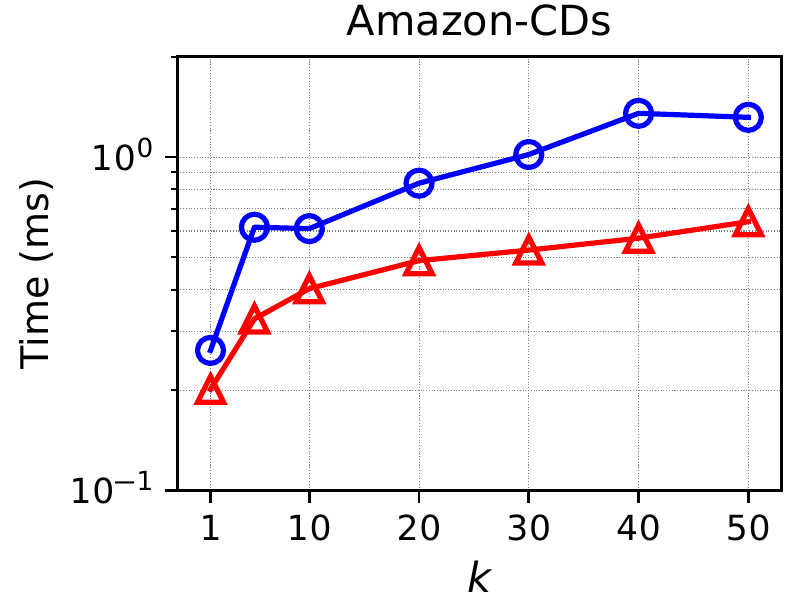}
  }
  \subcaptionbox{}[0.195\textwidth]{
    \includegraphics[height=1in]{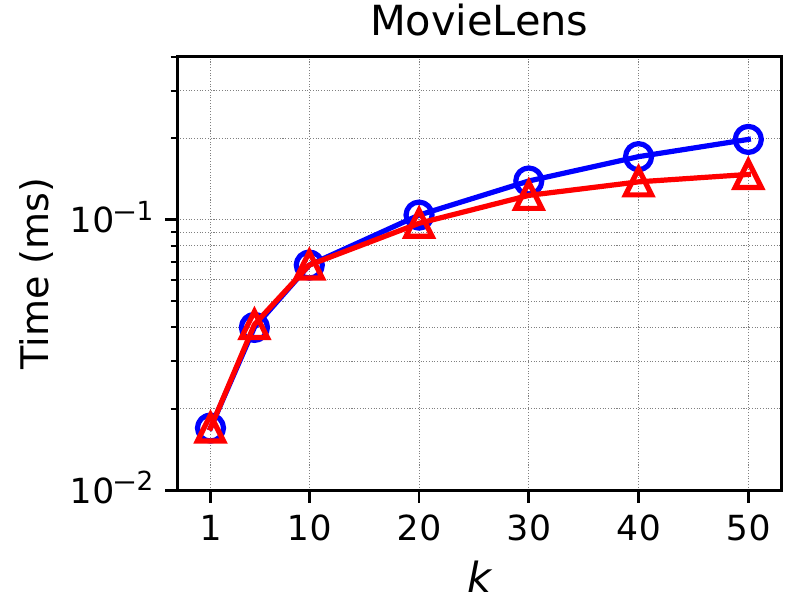}
  }
  \subcaptionbox{}[0.195\textwidth]{
    \includegraphics[height=1in]{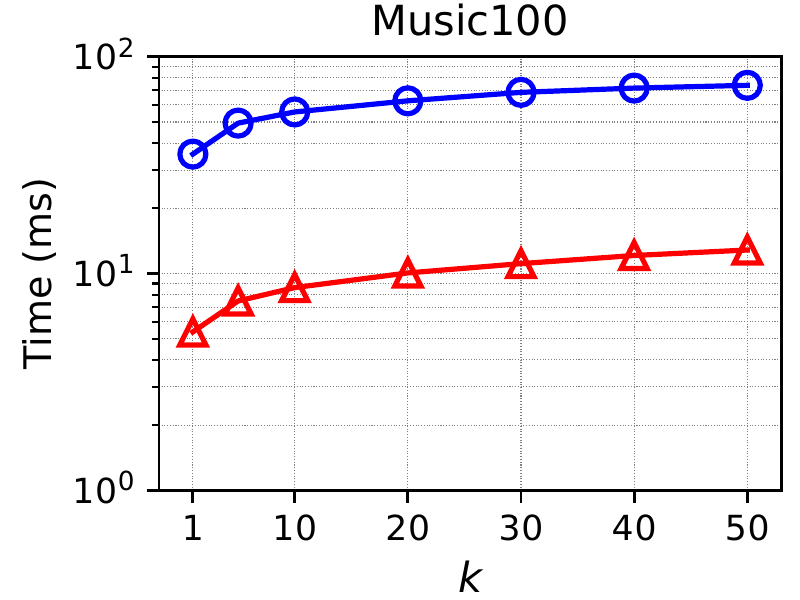}
  }
  \subcaptionbox{}[0.195\textwidth]{
    \includegraphics[height=1in]{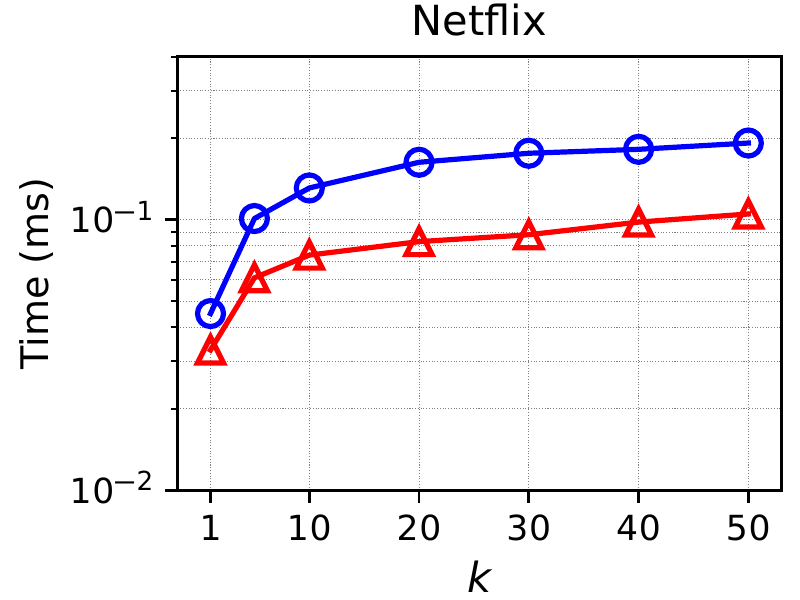}
  }
  \\\vspace{-1em}
  \subcaptionbox{}[0.195\textwidth]{
    \includegraphics[height=1in]{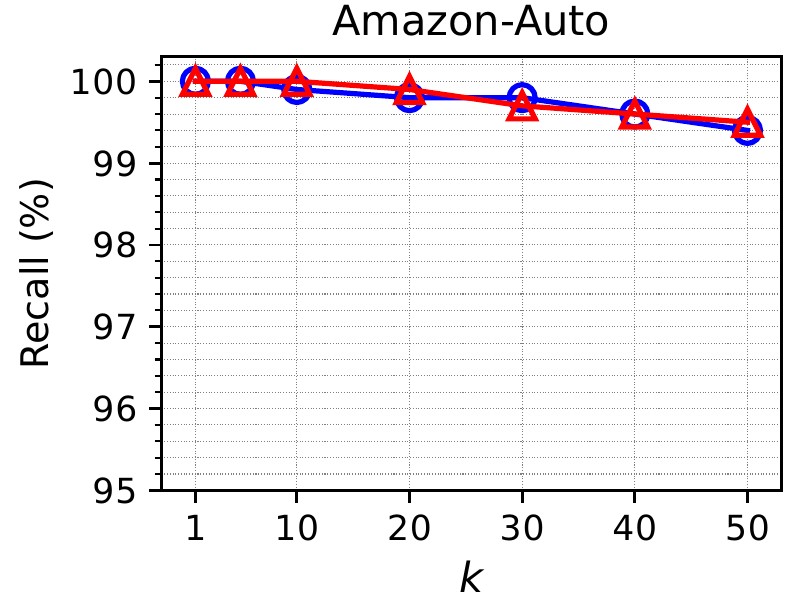}
  }
  \subcaptionbox{}[0.195\textwidth]{
    \includegraphics[height=1in]{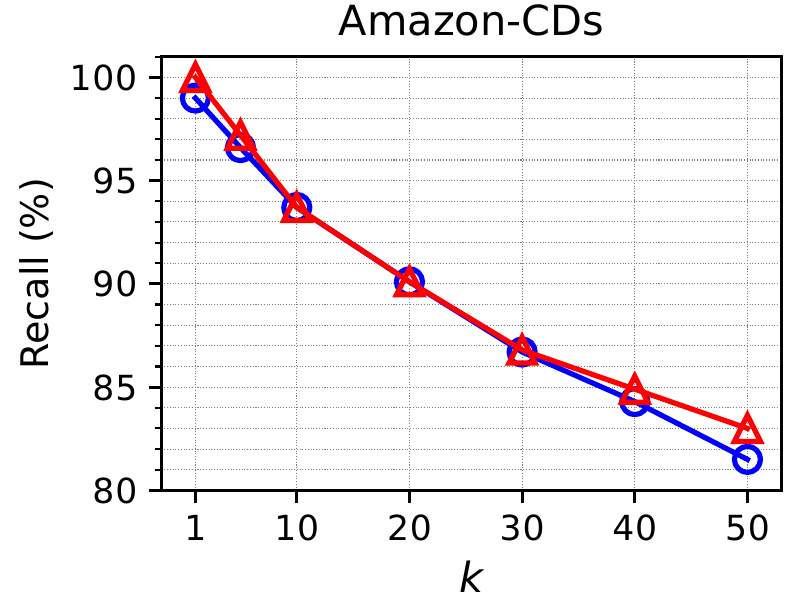}
  }
  \subcaptionbox{}[0.195\textwidth]{
    \includegraphics[height=1in]{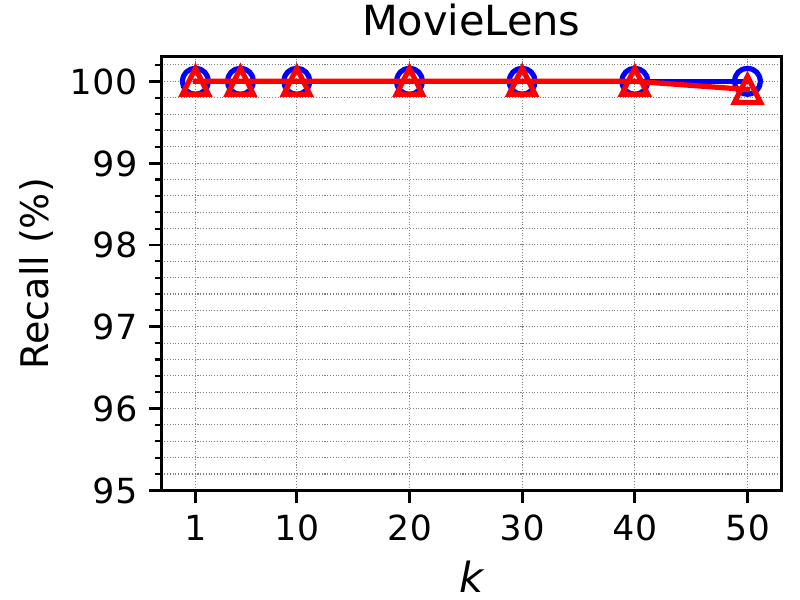}
  }
  \subcaptionbox{}[0.195\textwidth]{
    \includegraphics[height=1in]{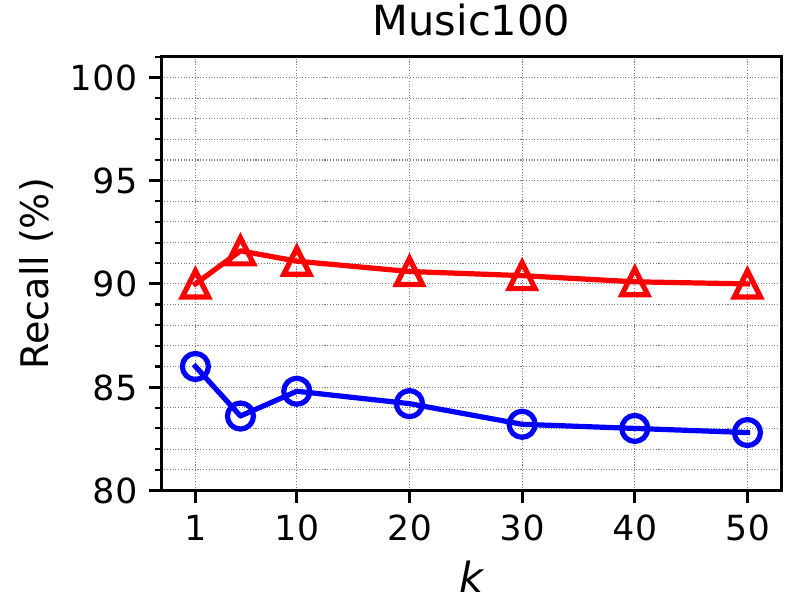}
  }
  \subcaptionbox{}[0.195\textwidth]{
    \includegraphics[height=1in]{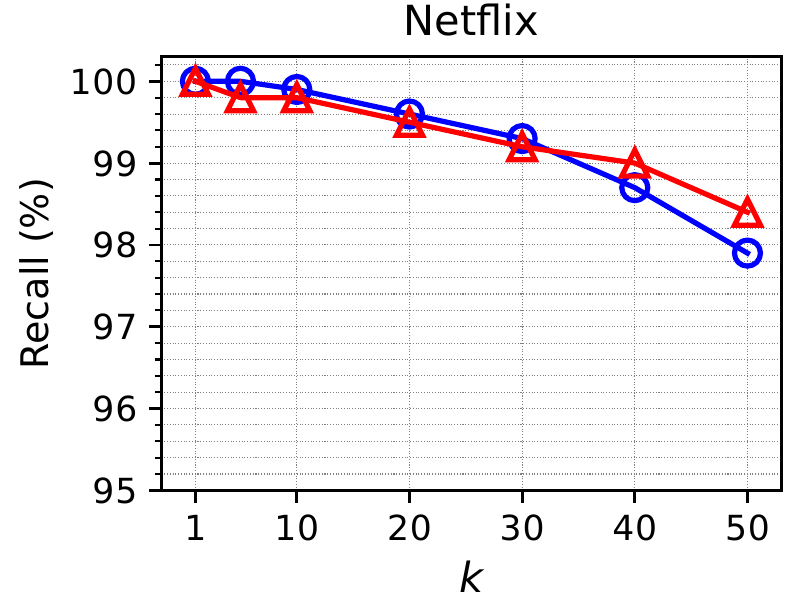}
  }
  \\\vspace{-1em}
  \subcaptionbox{}[0.195\textwidth]{
    \includegraphics[height=1in]{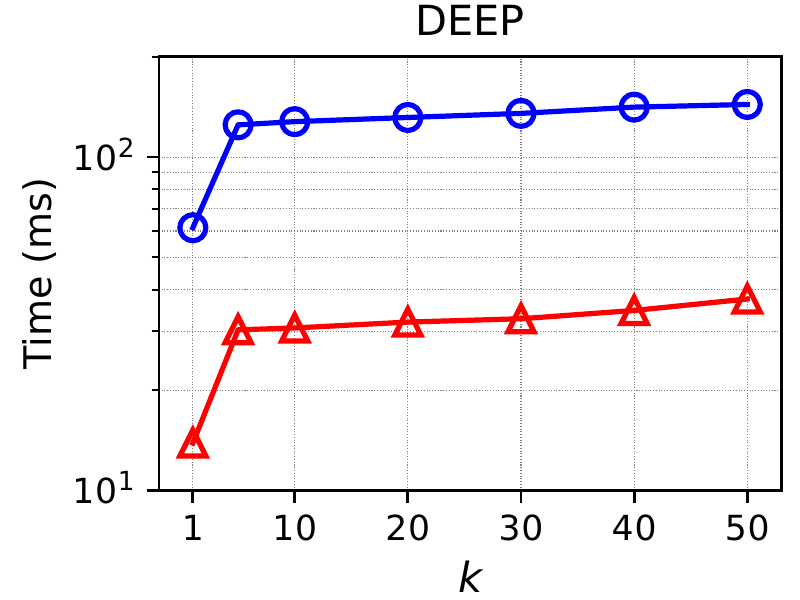}
  }
  \subcaptionbox{}[0.195\textwidth]{
    \includegraphics[height=1in]{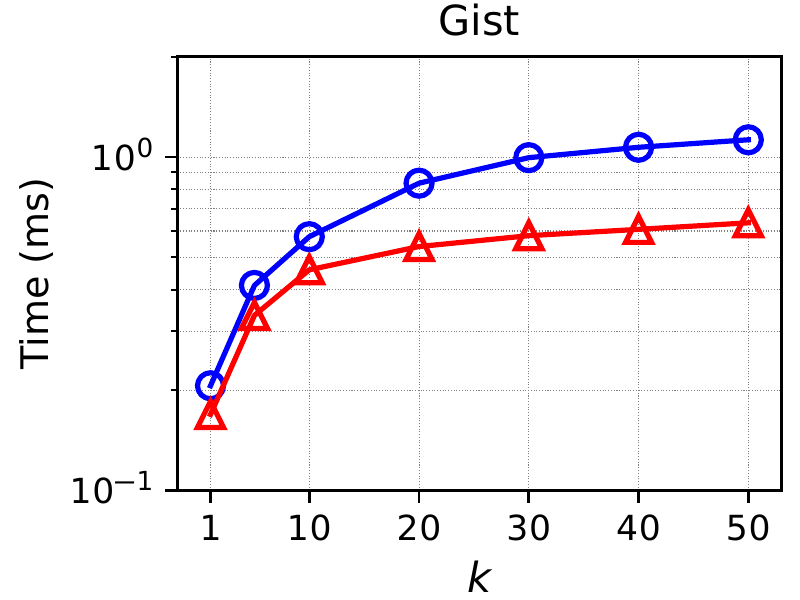}
  }
  \subcaptionbox{}[0.195\textwidth]{
    \includegraphics[height=1in]{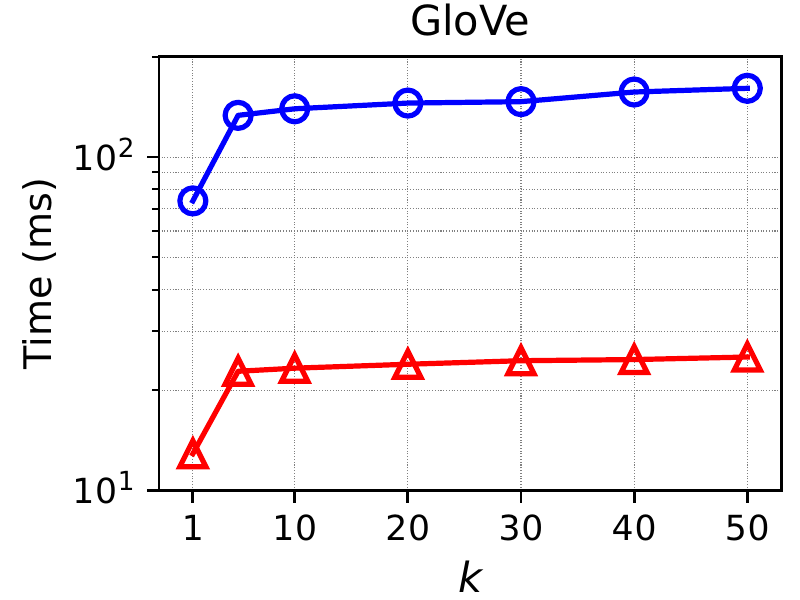}
  }
  \subcaptionbox{}[0.195\textwidth]{
    \includegraphics[height=1in]{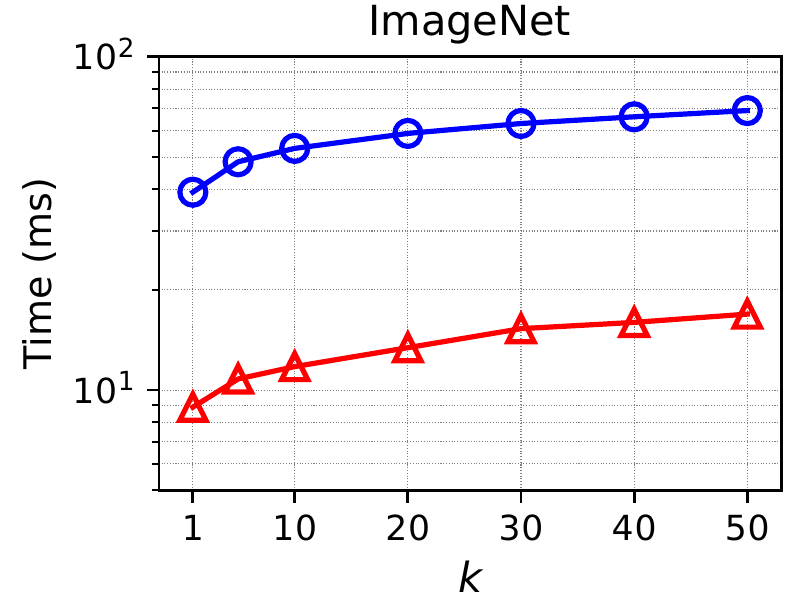}
  }
  \subcaptionbox{}[0.195\textwidth]{
    \includegraphics[height=1in]{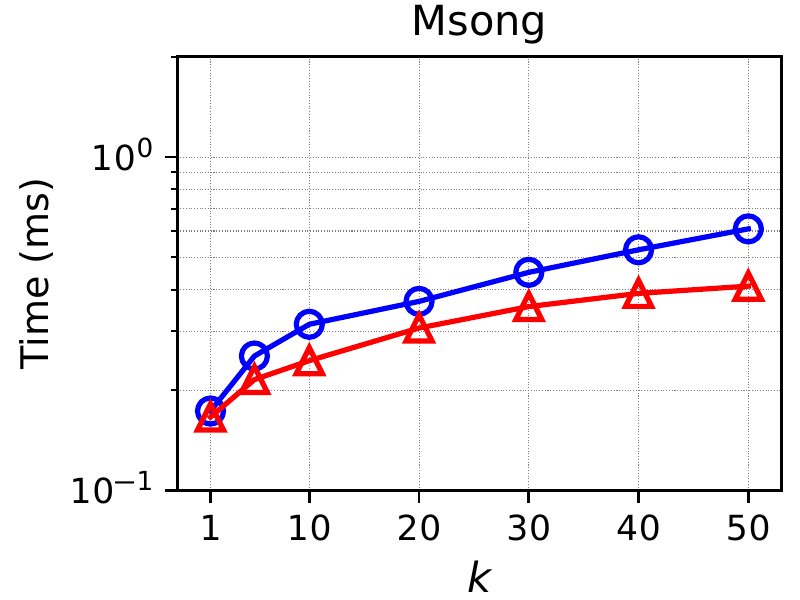}
  }
  \\\vspace{-1em}
  \subcaptionbox{}[0.195\textwidth]{
    \includegraphics[height=1in]{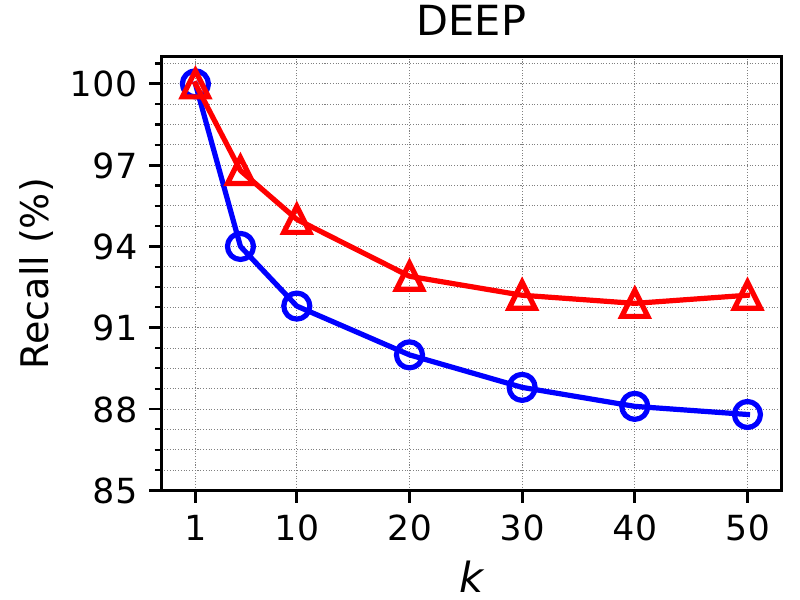}
  }
  \subcaptionbox{}[0.195\textwidth]{
    \includegraphics[height=1in]{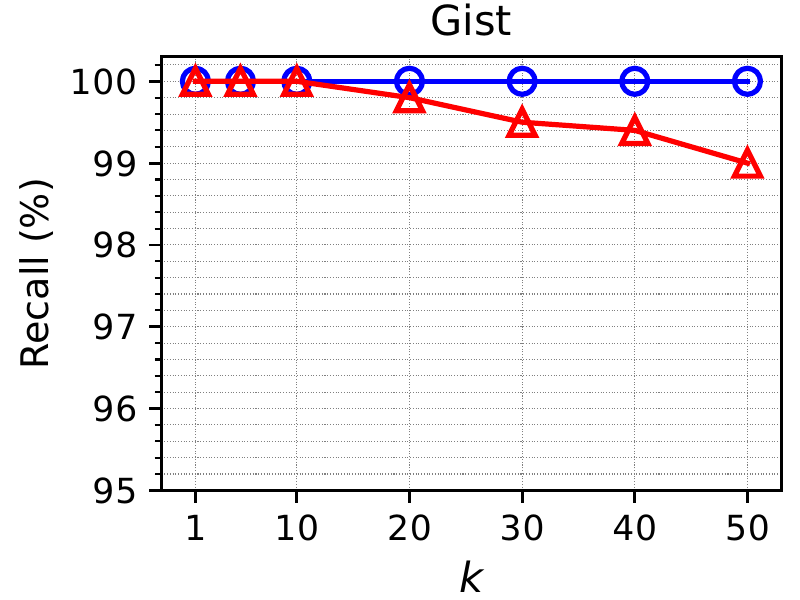}
  }
  \subcaptionbox{}[0.195\textwidth]{
    \includegraphics[height=1in]{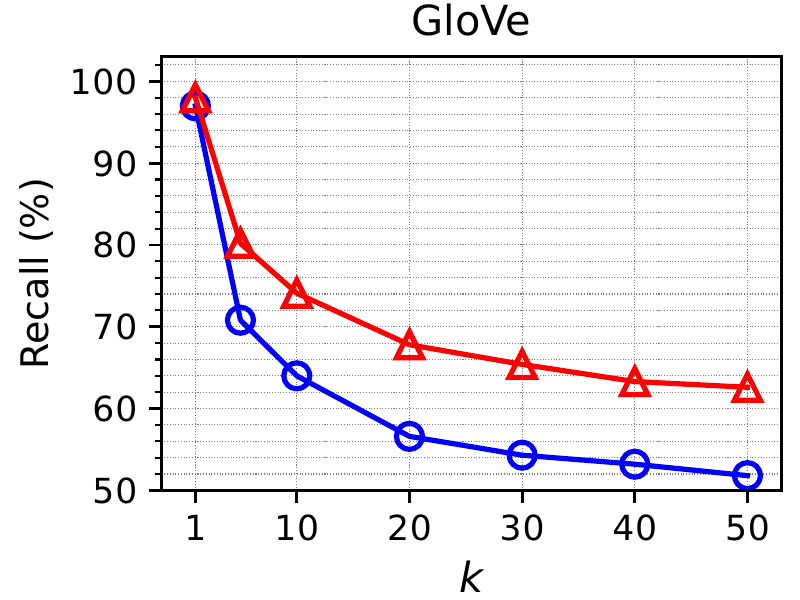}
  }
  \subcaptionbox{}[0.195\textwidth]{
    \includegraphics[height=1in]{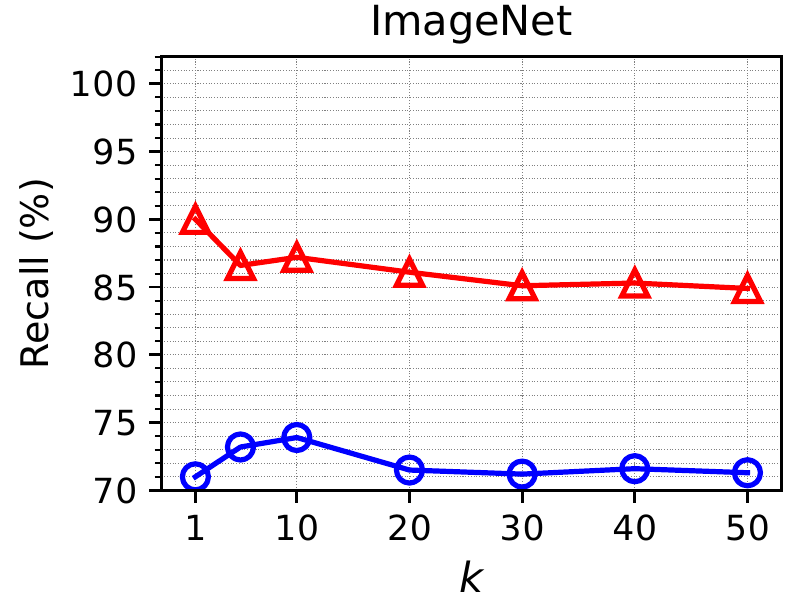}
  }
  \subcaptionbox{}[0.195\textwidth]{
    \includegraphics[height=1in]{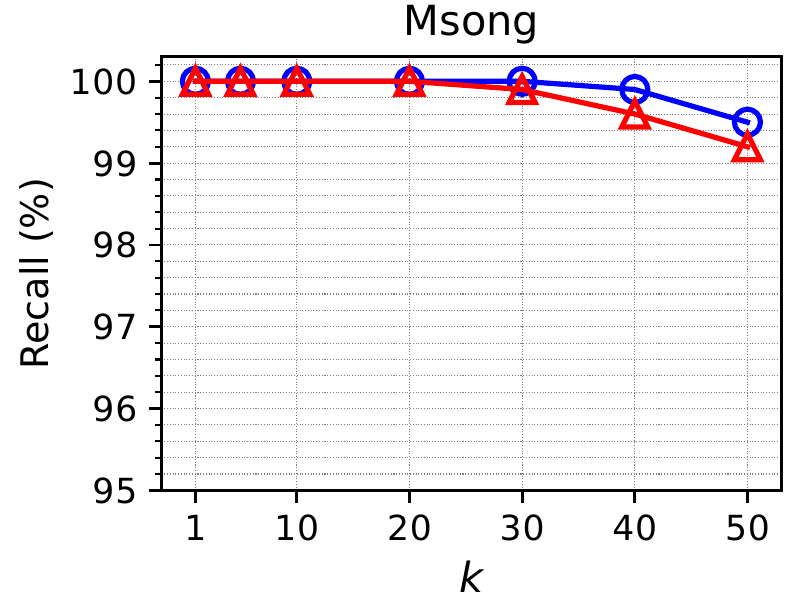}
  }
  \caption{Performance of H2-ALSH and SA-ALSH for $k$MIPS with $k \in \{1, 5, 10, 20, 30, 40, 50\}$.}
  \label{fig:exp:kmips}
  \vspace{-1.0em}
\end{figure*}

\paragraph{Impact of the Maximum Leaf Size $N_0$ for SAH.}
We then show how the maximum leaf size $N_0$ in the Cone-Tree affects the performance of SAH.
We fix $b=0.5$ and $K = 128$ and vary $N_0 \in \{20, 50, 100, 200\}$.
The results of SAH with $k=10$ are depicted in Figure~\ref{fig:exp:N_0}. Similar trends can be observed for other values of $k$.

From the bars of Figure~\ref{fig:exp:N_0}, we find that in most cases (except $N_0=20$ on the Amazon-Auto dataset), the query time of SAH with different values of $N_0$ varies very slightly.
Similar trends can be observed from the F1-scores of SAH except for the Amazon-CDs dataset. Thus, the impact of $N_0$ for SAH is not significant. By default, we follow~\cite{DBLP:conf/kdd/RamG12} to use $N_0 = 20$.

\paragraph{SA-ALSH vs. H2-ALSH for $k$MIPS.}
To verify our claim on the advantage of SA-ALSH over H2-ALSH, we compare their F1-scores and query time for $k$MIPS with varying $k \in \{1, 5, 10, 20, 30, 40, 50\}$.
In addition to the five recommendation datasets used in previous experiments, we also consider five real-world datasets that have been commonly used to benchmark $k$MIPS algorithms, namely DEEP,\footnote{\url{https://github.com/DBAIWangGroup/nns_benchmark/tree/master/data}} Gist,\footnote{\url{http://corpus-texmex.irisa.fr/}} GloVe,\footnote{\url{https://nlp.stanford.edu/projects/glove/}} ImageNet,\footnote{\url{https://github.com/DBAIWangGroup/nns_benchmark/tree/master/data}} and Msong.\footnote{\url{http://www.ifs.tuwien.ac.at/mir/msd/download.html}}
The number of vectors and the dimensionality $(n,d)$ of DEEP, Gist, GloVe, ImageNet, and Msong are ($1,000,000$, $256$), ($1,000,000$, $960$), ($1,183,514$, $100$), ($2,340,373$, $150$), and ($992,272$, $420$), respectively.
Note that these datasets cannot be used for R$k$MIPS as there is no user vector within them.

For the five recommendation datasets used in previous experiments, we use item vectors as the dataset and choose $100$ user vectors uniformly at random as queries; 
for the five new datasets, we randomly draw $100$ data vectors and remove them from the dataset as queries.
We tune the parameters in H2-ALSH and SA-ALSH so that they strike the best balance between accuracy and efficiency.
The F1-scores and query time of H2-ALSH and SA-ALSH for $k$MIPS with $k \in \{1, 5, 10, 20, 30, 40, 50\}$ are shown in Figure~\ref{fig:exp:kmips}.

We observe that SA-ALSH consistently outperforms H2-ALSH in terms of both query efficiency and result accuracy across all ten datasets.
Especially, SA-ALSH achieves up to $6.6\times$ speedups over H2-ALSH while acquiring at most $19\%$ higher F1-scores.
As the primary difference between SA-ALSH and H2-ALSH is the transformation scheme and other steps are mostly the same, these results directly confirm that SA-ALSH has lower distortion errors than H2-ALSH in the transformation.

\begin{table}[t]
\captionsetup{skip=3pt}
\centering
\caption{F1-scores ($\%$) of using the results of $k$MIPS to answer R$k$MIPS.}
\label{tbl:kmips:f1}
\resizebox{\columnwidth}{!}{%
\begin{tabular}{cccccccc}
\toprule
\textbf{Dataset} & $k=1$ & $k=5$ & $k=10$ & $k=30$ & $k=50$ \\ \midrule
Amazon-Auto & 0.0  & 0.0  & \textbf{0.1}  & \textbf{0.1}  & \textbf{0.1}  \\
Amazon-CDs  & 0.2  & 0.4  & 0.7  & 1.3  & \textbf{2.8}  \\
MovieLens   & 0.0  & \textbf{26.9} & 25.9 & 10.5 & 10.3 \\
Music100    & 11.1 & \textbf{31.7} & 30.6 & 21.7 & 18.6 \\
Netflix     & 5.1  & 5.7  & \textbf{7.5}  & 5.3  & 5.2  \\
\bottomrule
\end{tabular}
}
\end{table}

\paragraph{Difference between $k$MIPS and R$k$MIPS.}
To justify the difference between $k$MIPS and R$k$MIPS, we study the problem of whether the result of a $k$MIPS is a promising result of an R$k$MIPS for the same query vector.
We take the same set of item vectors used in previous experiments as query vectors for $k$MIPS and perform a linear scan to find the top-$k$ user vectors whose inner products with the query (item) vector are the largest.
Then, we measure the F1-scores of using the results of $k$MIPS to answer R$k$MIPS and show results in Table~\ref{tbl:kmips:f1}.

Since the F1-scores are at most $31.7\%$ and mostly below $10\%$, we confirm that the results of $k$MIPS and R$k$MIPS for the same query vector are not highly overlapped with each other and one cannot use the $k$MIPS result to answer the R$k$MIPS query.
This empirical evidence quantifies the differences in the applications of R$k$MIPS and $k$MIPS as well as verifies our claim  that \emph{the $k$MIPS might not be beneficial in finding potential customers for the query item}.

\end{document}